\documentclass[reprint, showkeys, 
nofootinbib,amsmath,amssymb, aps, prd,
]{revtex4-2}

\usepackage{graphicx}
\usepackage{dcolumn}
\usepackage{bm}
\usepackage{hyperref}

\begin{document}

\preprint{2201.10554}

\title{Novel aspects of integrability for NLSMs in symmetric spaces}
\author{Dimitrios Katsinis}
\email{dkatsinis@phys.uoa.gr}
\affiliation{Instituto de F\'isica, Universidade de S\~ao Paulo, Rua do Mat\~ao Travessa 1371, 05508-090 S\~ao Paulo, SP, Brazil}

%
%

\date{\today}

\begin{abstract}
We obtained the formal solution of the auxiliary system of Non Linear Sigma Models, whose target space is a rank 1 symmetric space based on the indefinite orthogonal group $O(p,q)$, corresponding to an arbitrary solution of the NLSM. This class includes Anti-de Sitter, de Sitter and Hyperbolic spaces, which are of interest in view of the AdS/CFT correspondence. The formal solution is related to the Pohlmeyer reduction of the NLSM, constituting another link between the NLSM and the reduced theory. Besides deriving the solution, we also review the Pohlmeyer reduction of such models. Finally, we comment on the implications for the monodromy matrix and its eigenvalues.
\end{abstract}

\maketitle


\section{Introduction}
\label{sec:introduction}

In the last 50 years there has been a tremendous number of studies, which point out that integrability is intertwined with high energy physics. This work concerns the integrability of Non Linear Sigma Models (NLSMs) on symmetric spaces. The general framework related to 2-dimensional integrable models was laid down in the 1970s and is still very important even for ongoing research. 

A particular class of such models is the Principal Chiral Model (PCM) \cite{Zakharov:1973pp}. The equations of motion of the PCM along with the flatness of the current are reproduced by the Lax connection
\begin{equation}\label{eq:Lax}
\mathcal{L}_\pm=\frac{\left(\partial_\pm g\right)g^{-1}}{1\pm\lambda},
\end{equation}
where $\lambda\in\mathbb{C}$ is the spectral parameter and $g$ is in general an element of some coset. The flatness of the latter is equivalent to the compatibility condition of the so-called auxiliary system, also known as the fundamental linear problem, which reads
\begin{equation}\label{eq:auxiliary_system_lax}
\partial_\pm \Psi(\lambda)=\mathcal{L}_\pm\Psi(\lambda).
\end{equation}

Classical integrability amounts to the existence of an infinite tower of conserved charges. This tower can be constructed for any Lax connection using the solution of the auxiliary system \eqref{eq:auxiliary_system_lax}. Defining the monodromy matrix $T(\sigma_f,\sigma_i;\lambda)$ as
\begin{equation}\label{eq:monodromy_matrix}
T(\sigma_f,\sigma_i;\lambda)=\Psi(\tau,\sigma_f;\lambda)\Psi^{-1}(\tau,\sigma_i;\lambda),
\end{equation}
it follows that its time derivative reads
\begin{multline}
\partial_\tau T(\sigma_f,\sigma_i;\lambda)= \mathcal{L}_\tau(\tau,\sigma_f)T(\sigma_f,\sigma_i;\lambda)\\-T(\sigma_f,\sigma_i;\lambda)\mathcal{L}_\tau(\tau,\sigma_i),
\end{multline}
where $\mathcal{L}_\tau$ is the component of the Lax connection when it is expressed in terms of the world-sheet coordinates $\sigma$ and $\tau$. In this case the relevant equation of the auxiliary system is $\partial_\tau \Psi(\tau,\sigma;\lambda)=\mathcal{L}_\tau\Psi(\tau,\sigma;\lambda)$. The constants $\sigma_i$ and $\sigma_f$ are determined by the specific solution and its boundary conditions. Depending on the latter, it is either the trace of the monodromy matrix that is conserved, which is the case for periodic boundary conditions, or the matrix itself if $\mathcal{L}_\tau$ vanishes both at $\sigma_i$ and $\sigma_f$. For open string boundary conditions one has to define a boundary monodromy matrix \cite{Cherednik:1984vvp,Sklyanin:1988yz}, which is again related to the solution of the auxiliary system. Expanding the monodromy matrix or its trace with respect to the spectral parameter one obtains the aforementioned infinite tower of conserved charges. The existence of all these conserved charges is such a strong constraint even on the quantized theory so that no particle production is allowed and the S-matrix of the theory is factorized into two particle S-matrices \cite{Luscher:1977uq}.

There is a far wider class of theories, which obviously includes the PCM, the so-called NLSMs. The action of such theories reads
\begin{equation}
S=\int d^2\sigma\, G_{\mu\nu}(X)\partial_+ X^\mu\partial_- X^\nu
\end{equation}
and their fields, i.e.  $X^\mu$, are mappings from the 2-dimensional world-sheet to a manifold $\mathcal{M}$. It turns out that if $\mathcal{M}$ is a symmetric space, the NLSM is integrable \cite{Eichenherr:1979ci,Eichenherr:1979hz}.\footnote{We remind the reader that a symmetric space $\mathcal{M}=F/G$ is defined as follows. Consider a group $F$ having a subgroup $G$, which correspond to the Lie algebras $\mathfrak{f}$ and $\mathfrak{g}$ respectively, so that $\mathfrak{f}$ admits the canonical decomposition $\mathfrak{f}=\mathfrak{g}\oplus\mathfrak{p}$. Then, $\mathcal{M}$ is a symmetric space if the following  commutation relations hold:
\begin{equation}
[\mathfrak{g},\mathfrak{g}]\subset\mathfrak{g},\qquad [\mathfrak{g},\mathfrak{p}]\subset\mathfrak{p},\qquad [\mathfrak{p},\mathfrak{p}]\subset\mathfrak{g}.
\end{equation}
} For further information about the integrability of NLSMs see \cite{Zarembo:2017muf,Driezen:2021cpd} and references therein. Finally, let us mention that integrability is not restricted to symmetric spaces. Recently there has been a lot of activity in the construction of integrable NLSMs \cite{Klimcik:2008eq,Delduc:2013fga,Delduc:2013qra,Sfetsos:2013wia,Klimcik:2014bta}, see also \cite{Klimcik:2021bjy,Hoare:2021dix} and references therein.

An interesting aspect of NLSMs on symmetric spaces is that besides being integrable, they are also reducible to integrable systems, which are multicomponent generalizations of the sine-Gordon equation. The original examples are the $O(3)$ and $O(4)$ NLSMs, corresponding to the spheres S$^2$ and S$^3$, which are related to the sine-Gordon  and the complex sine-Gordon equations, respectively, via the Pohlmeyer reduction \cite{Pohlmeyer:1975nb,Lund:1976ze,Lund:1976xd}. Subsequently many more symmetric spaces were studied including the N-sphere S$^{N}$, i.e. the $O(N+1)$ model \cite{Pohlmeyer:1979ch}, and $\mathbb{CP}^N$ \cite{Eichenherr:1979uk}, as well as (Anti) de Sitter space \cite{Barbashov:1980kz,DeVega:1992xc,Larsen:1996gn}.

Essentially, the reduction amounts to introducing some degrees of freedom, which are connected non-locally to the fields of the NLSM, and setting the non-vanishing components of the stress-energy tensor to constants. This is always possible as the NLSM is conformally invariant at the classical level. When the symmetric space $\mathcal{M}$ is of positive definite signature, which, for instance, is the case for S$^N$ and $\mathbb{CP}^N$, it follows that the constants are necessarily positive, thus $T_{\pm\pm}$ read
\begin{equation}\label{eq:stress_tensor}
T_{\pm\pm}=G_{\mu\nu}\partial_{\pm} X^\mu\partial_{\pm}X^\nu=m_\pm^2,
\end{equation}
in appropriate coordinates. One should keep in mind that the Pohlmeyer reduction is a many-to-one mapping. In particular, the reduced theory depends on $m_+$ and $m_-$ only via their product $m_+m_-$ and not on their ratio. We are going to refine this statement, but it is sufficiently accurate for now. Evidently, even though integrability is preserved, for $m_+m_-\neq0$ the conformal invariance is broken by the reduction. Finally, it is worth pointing out that when the action of the NLSM on the symmetric space is equivalent to free fields on a higher-dimensional space subject to a quadratic constraint, the equations of motion of the NLSM become linear for a given solution of the reduced theory.

Using a group theoretical approach, the reducibility of the NLSM on symmetric spaces was established in \cite{DAuria:1979ham,DAuria:1980iyh,DAuria:1980ucx}. This approach is extremely powerful leading to many interesting developments. These reduced theories are known as Symmetric Space sine-Gordon models (SSSGs). For a long time their Langrangian formulation was an open problem, until it was solved for symmetric spaces of rank 1 in \cite{Bakas:1995bm}, see also \cite{Park:1994bx,Hollowood:1994vx,Fernandez-Pousa:1996aoa}.\footnote{Recall that the rank of a symmetric space $F/G$ is the dimension of the maximal Abelian subspaces in the orthogonal complement of $\mathfrak{g}$ in $\mathfrak{f}$.} Specifically, it was shown that SSSGs are gauged Wess-Zumino-Witten (WZW) models models perturbed by an appropriate potential, which preserves integrability. For a recent review on the group theoretical approach of Pohlmeyer reduction, as well as for an exhaustive analysis of the Lagrangian formulation of SSSGs, see \cite{Miramontes:2008wt}.

Another intriguing characteristic of integrable NLSMs regards the construction of new solutions when a solution is already known. This is achieved with the application of the dressing method \cite{Zakharov:1973pp,Zakharov:1980ty,Harnad:1983we}. In order to do so one has to solve the auxiliary system \eqref{eq:auxiliary_system_lax}, where $g$ corresponds to the known NLSM solution, which in this context is referred to as the seed solution. Given the solution of the auxiliary system one can systematically construct an infinite tower of new NLSM solutions. 

There is an analogous story for the reduced systems. There are the so-called B\"acklund transformations for these systems, which are sets of first order non-linear equations and allow the derivation of new solutions given a known one. In the case of the sine-Gordon equation these transformations essentially insert solitons on the background of the known solution. Multiple solutions corresponding to the same known solution can be combined using addition formulas, for example see \cite{Park:1995np}. The B\"acklund transformations are the counterpart of the dressing transformations. In particular, it has been shown that a dressing transformation of the NLSM solution, automatically performs a B\"acklund transformation to its avatar in the reduced theory \cite{Hollowood:2009tw}. It is also important that the B\"acklund transformations generate an infinite tower of conserved charges \cite{Pohlmeyer:1975nb}. These charges are related to the ones constructed via the expansion of the monodromy matrix \cite{Arutyunov:2003rg,Arutyunov:2005nk}.

In view of the AdS/CFT correspondence \cite{Maldacena:1997re,Gubser:1998bc,Witten:1998qj}, which relates planar strongly coupled $\mathcal{N}=4$ Super-Yang-Mills to classical free IIB string theory in AdS$_5\times$S$^5$, the interest on the reduced models was revived. In this setup the dynamics of the superstrings is determined by the Metsaev-Tseytlin action \cite{Metsaev:1998it}. The theory is formulated as the super-coset $\frac{PSU(2,2\vert4)}{SO(1,4)\times SO(5)}$ and is classically  integrable \cite{Bena:2003wd}. As this coset is a symmetric space having a $\mathbb{Z}_4$ grading, the theory is reducible and the corresponding models were constructed in \cite{Grigoriev:2007bu,Mikhailov:2007xr,Grigoriev:2008jq,Hoare:2012nx}. Moreover the reduced theory is finite in the UV \cite{Roiban:2009vh}  and the properties of the classical charges indicate that the theory is supersymmetric \cite{Schmidtt:2011gtt,Schmidtt:2011nr}. The world-sheet supersymmetry is non-local \cite{Goykhman:2011mq,Hollowood:2011fq}. Similarly, the bosonic reduced model corresponding to AdS$_4\times\mathbb{CP}^3$, which is dual to the Aharony-Bergman-Jafferis-Maldacena (ABJM) theory \cite{Aharony:2008ug} and is formulated as the $\frac{OSP(2,2\vert6)}{SO(1,3)\times U(3)}$ super-coset \cite{Arutyunov:2008if,Stefanski:2008ik}, was constructed in \cite{Rashkov:2008rm}. The supersymmetric one was presented in \cite{Dukalski:2009pr}. The reduced models, maintaining 2-dimensional Lorentz invariance, provide an alternative approach to the standard light-cone description of the on-shell degrees of freedom.

Let us be more specific about the reduced models in this context. Equation \eqref{eq:stress_tensor} may be interpreted as the Virasoro constraint for strings propagating in $\mathbb{R}_t\times\mathcal{M}$, where a linear gauge for the time coordinate $t$ of the target space has been employed, namely 
\begin{equation}
t=m_+\sigma^++m_-\sigma^-.
\end{equation}
This gauge, which is a generalization of the usual static one, i.e. $t=m\tau$, facilitates the rest of our work. A world-sheet boost with appropriate velocity turns the linear gauge to the static one. At the level of equations \eqref{eq:stress_tensor} setting $m_\pm\rightarrow c^{\pm1} m_\pm$, the parameter $c$ can be absorbed via a boosted form of the embedding functions, i.e. by $X^\mu(\sigma^+,\sigma^-;m_+,m_-)\rightarrow X^\mu(c\, \sigma^+,c^{-1} \sigma^-;m_+,m_-)$. Notice that the product $m_+m_-$ remains invariant. Of course the counterpart in the reduced theory depends on the boost too. As the boost to the static gauge depends also on the ratio of $m_+$ and $m_-$ the Pohlmeyer avatar no longer depends solely on the product $m_+m_-$. So, to be more precise the reduced theory depends only on this product in an appropriate frame.

Many properties of classical string solutions are captured by the reduced models. Strings, whose world-sheet is infinite, correspond to solitonic solutions of the reduced system. For instance, the giant magnon \cite{Hofman:2006xt} ($\mathbb{R}_t\times\textrm{S}^2$), the dyonic giant magnon \cite{Chen:2006gea} ($\mathbb{R}_t\times\textrm{S}^3$) and the single spike \cite{Ishizeki:2007we} ($\mathbb{R}_t\times\textrm{S}^2$ and $\mathbb{R}_t\times\textrm{S}^3$) correspond to solitonic solutions of the sine-Gordon and complex sine-Gordon equations. In an analogous way, the BMN particle \cite{Berenstein:2002jq}, which moves at the speed of light on the equator of S$^5$, corresponds to the (stable) vacuum solution of the sine-Gordon equation. Similar identifications exist for for kink-train solutions of the sine-Gordon and complex sine-Gordon equation \cite{Okamura:2006zv,Hayashi:2007bq}. See also \cite{Katsinis:2018zxi} for a complete classification of elliptic string solutions in $\mathbb{R}_t\times\textrm{S}^2$. Similar conclusions also hold for strings in AdS and dS \cite{DeVega:1992xc,Larsen:1996gn,Jevicki:2007aa,Jevicki:2008mm,Bakas:2016jxp}, as well as for dressed elliptic strings in $\mathbb{R}_t\times\textrm{S}^2$ \cite{Katsinis:2018ewd,Katsinis:2019oox}.

Besides the relation between the string solutions and their counterparts in the reduced theory, AdS/CFT correspondence is intimately related to integrability in many more aspects, such as the spectrum of dual theories and the calculation of correlation functions and Wilson loops, see \cite{Beisert:2010jr} for a review. Of particular interest is the spectral problem. Today there are methods that determine the (quantum mechanically) exact spectrum of $\mathcal{N}=4$ Super-Yang-Mills, see \cite{Gromov:2017blm} and references therein, but making contact with them is way beyond the scope of this work. The NLSM and all the aspects of integrability that we mentioned, i.e. the Pohlmeyer reduction, dressing method, B\"acklund transformations, regard only the classical geometry. In a series of papers \cite{Kazakov:2004qf,Kazakov:2004nh,Beisert:2004ag,Beisert:2005bm,Beisert:2005di} it was shown that the single trace operators of planar $\mathcal{N}=4$ Super-Yang-Mills in the thermodynamic limit, i.e. in the case of infinitely many insertion of fields, and the NLSM on AdS$_5\times$S$^5$ share a spectral curve.\footnote{As a side note let us mention that the general solution of the NLSMs in spaces of constant curvature is expressed in terms of hyperelliptic theta and related functions. These functions are defined in terms of a spectral curve, which determines their periodicities. Essentially this construction, known as finite gap integration, generalizes the mode expansion of flat space to an expansion in terms of the genus of the spectral curve in the case of curved spaces, see the introduction of \cite{Beisert:2005bm}. Using either the spectral curve directly, or the general solution of the Pohlmeyer reduced theory expressed in terms of theta functions along with properties of these functions, one can construct the general solution of the NLSM. Nevertheless, we should point out that explicit solutions are known only in the case of the $\mathbb{R}\times\textrm{S}^3$ NLSM \cite{Dorey:2006zj} (using the first method) or the Euclidean NLSM on $\textrm{H}^3$ (using the second method) \cite{Ishizeki:2011bf,Kruczenski:2013bsa}.} Similarly, the spectral curve corresponding to the super-coset $\frac{OSP(2,2\vert6)}{SO(1,3)\times U(3)}$ was constructed in \cite{Gromov:2008bz}. We remind the reader that the spectral curve is directly related to the eigenvalues of the monodromy matrix $T$. This result establishes that both dual theories, in this specific limit, share their conserved charges. However, even though this result is powerful, the identification of a specific string solution, which is dual to a particular operator is not straightforward. A systematic approach for this problems is elusive.

In the framework of AdS/CFT, the expectation values of Wilson loops at strong coupling are calculated by the area of minimal surfaces whose boundary is the loop in question \cite{Maldacena:1998im,Rey:1998ik}. Interestingly enough, null polygonal Wilson loops are related to gluon scattering amplitudes at strong coupling \cite{Alday:2007hr,Alday:2007he}, see also \cite{Alday:2009yn,Alday:2009dv}. Moreover, by introducing an appropriate contour in the internal space, the Wilson loop can be made supersymmetric \cite{Zarembo:2002an,Drukker:2007dw}. Implementing  localization, the supersymmetric Wilson loops can by calculated without the use of holography, directly in field theory, exactly (sometimes even for finite $N$) \cite{Pestun:2007rz,Pestun:2016zxk}. The comparison of the results of both calculations is a highly  non-trivial test of AdS/CFT. Finally, minimal surfaces are also relevant for the calculation of Holographic Entanglement Entropy (HEE) \cite{Ryu:2006bv,Ryu:2006ef}. The HEE corresponding to a region defined by some entangling surface equals the area of co-dimension 2 minimal surface, which extends in the bulk and whose boundary is the entangling surface. In the case of AdS$_4$ the minimal surfaces are 2-dimensional, thus described by a NLSM. All the aforementioned space-like minimal surfaces in AdS or in the Hyperbolic space are described by NLSMs and also admit a Pohlmeyer reduction \cite{Dorn:2009kq,Dorn:2009gq,Jevicki:2009bv,Burrington:2009bh,Dorn:2009hs,Ishizeki:2011bf,Kruczenski:2013bsa,Kruczenski:2014bla,Pastras:2016vqu,He:2017cwd}.

The above analysis demonstrates that the auxiliary system \eqref{eq:auxiliary_system_lax} is significant for many reasons. In our previous works we applied the dressing method for elliptic strings in $\mathbb{R}_t\times\textrm{S}^2$ \cite{Katsinis:2018ewd}, see also \cite{Arutyunov:2003rg}, and elliptic minimal surfaces in H$^3$ \cite{Katsinis:2020dhe}. One could proceed in a systematic manner and the only equations that had to be solved, were essentially solved by the seed solution upon altering some parameters. Motivated by this observation in \cite{Katsinis:2020avk} we obtained the formal solution of the auxiliary system of the $O(3)$ NLSM corresponding to an \emph{arbitrary} seed. The solution is constructed by combining appropriately the seed solution with a ``virtual'' one. The latter is obtained from the seed solution by the substitution
\begin{equation}\label{eq:m_rescalling}
m_\pm\rightarrow\frac{1\mp\lambda}{1\pm\lambda}m_\pm,
\end{equation}
where $\lambda$ is the spectral parameter. Thus, the virtual solution solves the equation of motion of the NLSM, has the same avatar in the reduced theory as the seed one, but as  $\lambda\in\mathbb{C}$ it belongs to the complexification of the coset. We should also point out that the rescaling spoils the boundary conditions of the seed solution, which is important for our results. \footnote{Considering seed solutions satisfying periodic boundary conditions, the virtual solution is periodic for specific values of the spectral parameter. In the context of the dressing method this translates to specific locations of the poles of the so-called dressing factor. It turns out that dressed solutions corresponding to such poles are either stable or unstable perturbations of the seed solution. This is shown explicitly for dressed elliptic strings in $\mathbb{R}\times\textrm{S}^2$ \cite{Katsinis:2019oox}, but it is expected to be true for any seed solution.}  A direct consequence of this result is the existence of a non-linear superposition rule at least for the specific NLSM. Regarding the dressing method the concept of superposition is already there as one can construct an infinite tower of NLSM solution by solving the auxiliary system once. Nevertheless, it turns out that no differential equation has to be solved. The dressing method is the implementation of a non-linear superposition in the first place. At the level of the reduced theory this result implies that there is no need to solve the B\"acklund transformations and that solitons are inserted for free. These conclusions present a novel aspect of NLSMs integrability, which expands the framework built in the 1970s and later on. As the derivation of \cite{Katsinis:2020avk} is model specific, the generalization of these conclusions could be questionable and the structure of the superposition obscure.

In our recent work \cite{Katsinis:2022amo} we were able to obtain the formal solution the auxiliary system for the whole class of $O(n)$ NLSMs. The solution relies on the Pohlmeyer reduction. Even though we presented the group theoretical approach for the study of the reduced models, there exists a geometric one, which treats the reduction as an embedding problem, introduced in \cite{Lund:1976ze,Lund:1976xd}, see also \cite{Barbashov:1980kz,Barbashov:1983cs,Bakas:1996uw}. One considers the target space as a sub-manifold of an enhanced flat space and analyzes the embedding of the NLSM solution in the enhanced space. In order to do so one introduces a basis in the enhanced space. It turns out that the solution of the auxiliary system is related to this basis, as well as on the redefinition of the parameters $m_\pm$ given by \eqref{eq:m_rescalling}. Moreover, in the case of periodic boundary conditions, the monodromy matrix is non-trivial just because the virtual solution is not periodic, since the rescaling spoils the periodic boundary conditions.

The purpose of this work is twofold. In view of AdS/CFT correspondence the solution of the auxiliary system in the case of AdS space presents a lot of interest. Of course, it is appealing to find the formal solution of the auxiliary system for more symmetric spaces. To kill two birds with one stone, we study the rank 1 symmetric spaces corresponding to the indefinite orthogonal group $O(p,q)$.

The structure of the paper is as follows. In section \ref{sec:Pohlmeyer_Reduction} we perform the Pohlmeyer reduction of rank 1 symmetric spaces corresponding to the indefinite orthogonal group $O(p,q)$. In section \ref{sec:Auxiliary} we solve the auxiliary system. Finally, in section \ref{sec:charges} we analyze the structure of the monodromy matrix and as an indicative example we derive its eigenvalues in the case of elliptic strings in $\mathbb{R}_t\times\textrm{S}^2$. There are two appendices. In appendix \ref{sec:AdS} we provide additional information for the Pohlmeyer reduction of AdS. Appendix \ref{sec:details} contains some details relevant for the derivation of section \ref{sec:Auxiliary}.

\section{Pohlmeyer reduction}\label{sec:Pohlmeyer_Reduction}

It is well known that strings propagating in symmetric spaces such as AdS and dS are related to integrable models, namely sinh-Gordon or cosh-Gordon and multicomponent generalizations of them, via the Pohlmeyer reduction \cite{Pohlmeyer:1975nb,Lund:1976ze,Lund:1976xd}. There is extensive work on this subject in the literature \cite{Barbashov:1980kz,DeVega:1992xc,Larsen:1996gn}. In this section, we generalize these works, while introducing our own convention, which will facilitate the second part of the paper. Since the aforementioned spaces are of indefinite signature, one should distinguish between different types of reductions. The tangent vectors of the world-sheet can be space-like, light-like or time-like and consequently the $T_{\pm\pm}$ components of the stress tensor positive, negative or vanishing, see \eqref{eq:stress_tensor}. The reduction is characterized accordingly. 

In the context of string theory the various types of reduction correspond to different geometric setups. Light-like reduction can be interpreted as stings on the non-compact manifold $\mathcal{M}$. If the reduction is time-like the geometry is $\mathcal{M}\times\mathcal{M}_c$, where $\mathcal{M}_c$ is a compact manifold having positive definite metric. In the case of space-like reduction the theory is defined in $\mathcal{M}_{nc}\times\mathcal{M}$, where $\mathcal{M}_{nc}$ is another non-compact manifold. The case $\mathbb{R}_t\times\mathcal{M}$ analyzed in the introduction is a just an example. Finally, let us mention that the geometry of the last case admits a time-like reduction too. In the context of the AdS/CFT correspondence light-like and time-like reduction of non-compact spaces are more natural. Having analyzed all options, let us introduce our conventions. 

It is well known that $\textrm{AdS}_{n+1}$ and $\textrm{dS}_{n+1}$ are sub-manifolds of $\mathbb{R}^{(2,n)}$ and  $\mathbb{R}^{(1,n+1)}$, respectively. They are both symmetric spaces realized as the $SO(2,n)/SO(1,n)$ and $SO(1,n+1)/SO(1,n)$ cosets  respectively. We can deal with both spaces simultaneously by introducing the metric tensor $\eta_{\mu\nu}=\textrm{diag}(-1,1,\dots,1,s)$, where $s=-1$ corresponds to $\textrm{AdS}_{n+1}$ and $s=1$ to $\textrm{dS}_{n+1}$. Denoting the vectors of the enhanced space as $ Y^{\mu}$, where  $Y^{\mu}=(Y^{-1},Y^{0},Y^{1},\dots Y^{n})$, the spaces of interest correspond to the sub-manifolds
\begin{equation}\label{eq:geometric_constraint}
Y\cdot Y=s\Lambda^2.
\end{equation}
It goes without saying that the inner product is defined as $A\cdot B=\eta_{\mu\nu} A^{\mu}B^{\nu}$.

As the purpose of this work is to study the relation between the solution of the auxiliary system and the Pohlmeyer reduction beyond the n-sphere \cite{Katsinis:2022amo}, we will consider more general geometries. Rather than restricting our analysis only on AdS and dS spaces, we will go one step further and study rank 1 cosets based on the indefinite orthogonal group. In order to do so we introduce the metric
\begin{equation}\label{eq:metric_enhanced}
\eta_{\mu\nu}=\begin{pmatrix}
 I_{p,q} & 0\\
 0 & s
\end{pmatrix},\qquad I_{p,q}=\begin{pmatrix}
-I_p & 0\\
0 & I_q
\end{pmatrix},
\end{equation}
where $I_k$ denotes the $k\times k$ identity matrix and $p+q=n+1$. It the following we use the notation $I_{q,p}$ for matrices of this specific form. For $s=+1$ the space is a sub-manifold of $\mathbb{R}^{p,q+1}$, it has positive constant curvature and corresponds to the coset $SO(p,q+1)/SO(p,q)$. Similarly, for  $s=-1$ the space is a sub-manifold of $\mathbb{R}^{p+1,q}$, it has negative constant curvature and corresponds to the coset $SO(p+1,q)/SO(p,q)$. Of course both spaces are related by the the duality $s\rightarrow-s$ and $p\leftrightarrow q$ \cite{Helgason}. Finally we remind the reader that for $s=-1$, $p=0$ and $q=n+1$ we obtain the Hyperbolic space H$^{n+1}$.

\subsection{The NLSM action}
Having set the ground, we can discuss the NLSM. The corresponding action reads
\begin{equation}
S=\int d\sigma^+ d\sigma^-\left[\partial_+Y\cdot\partial_-Y+\lambda\left(Y\cdot Y-s\Lambda^2\right)\right],
\end{equation}
where $\lambda$ is a Lagrange multiplier. The fields $Y$ are vectors of the enhanced space $\mathbb{R}^{p,q+1}$ or $\mathbb{R}^{p+1,q}$, depending on the value of $s$, which are restricted on the sub-manifold \eqref{eq:geometric_constraint} via the Lagrange multiplier. The equations of motion read
\begin{equation}
\partial_+\partial_-Y=\lambda Y.
\end{equation}
It is straightforward to calculate the Lagrange multiplier, which is given by
\begin{equation}
\lambda=-\frac{s}{\Lambda^2}\partial_+Y\cdot\partial_-Y,
\end{equation}
implying that the equations of motion assume the form
\begin{equation}\label{eq:EOM_NLSM}
\partial_+\partial_-Y=-\frac{s}{\Lambda^2}\left(\partial_+Y\cdot\partial_-Y\right) Y.
\end{equation}
The equations of motion are accompanied by the conservation of the stress energy tensor. The precise form of the components of the latter depends on the theory under consideration and more specifically on whether one considers the target space consisting solely of the aforementioned manifolds, or if the target space is the direct product of these spaces with another manifold. For instance, in view of the AdS/CFT correspondence, its very usual to consider the target spaces $\textrm{AdS}_5\times\textrm{S}^5$ or $\textrm{AdS}_4\times \mathbb{CP}^3$. Keeping an open mind and adopting a mathematical perspective, we consider the non-vanishing components of the stress tensor to be
\begin{equation}\label{eq:Virasoro}
T_{\pm\pm}=\partial_\pm Y\cdot \partial_\pm Y= t m^2_\pm.
\end{equation}
The case $t=0$ gives rise to light-like reduction, while $t=1$ to space-like and $t=-1$ to time-light. In the last two cases, string theory interpretation requires the introduction of another manifold so that the Virasoro constraints corresponding to the whole target space are satisfied.
\subsection{The embedding problem}
In order to perform the Pohlmeyer reduction, we consider the embedding problem of the NLSM solution $Y$ in the enhanced space. To do so, we introduce a basis in the enhanced space. Of course, this basis includes the vectors $Y$, $\partial_\pm Y$, which span the tangent space of the string world-sheet, and $n-1$ linearly independent vectors. Essentially, we consider the enhanced space as the direct product of the 3-dimensional space $\mathcal{M}_3$, spanned by $Y$ and $\partial_\pm Y$ and the $(n-1)$-dimensional space $\mathcal{M}_{n-1}$, spanned some vectors $v_i$, where $i=1,\dots,(n-1)$.\footnote{Lower-case Latin letters run from 1 to $n-1$, while lower-case Greek letters run from 1 to $n+2$.} We normalize these vectors as
\begin{equation}
v_i\cdot v_j=s_i\delta_{i,j},
\end{equation}
where $s_i=\pm1$. The careful reader will realize that this definition is sloppy. For the moment let us comment that since the basis spans either $\mathbb{R}^{p,q+1}$ or $\mathbb{R}^{p+1,q}$ the total number of $s_i$ having the value $+1$ and of the ones having the value $-1$ is constrained. To make the notation more uniform, we also set
\begin{equation}
v_{n+2}=Y,\qquad v_{n+1}=\partial_+Y,\qquad v_{n}=\partial_-Y.
\end{equation}
Thus, we have the following basis $\mathcal{V}=\{v_1,\dots ,v_{n+2}\}.$
Considering the $\mathcal{M}_3$ subspace, the norm of $v_{n+2}$ is fixed by \eqref{eq:geometric_constraint}, while the norms of $v_{n}$ and $v_{n+1}$ are fixed by the equations \eqref{eq:Virasoro}. The geometric constraint also implies that $v_{n}$ and $v_{n+1}$ are perpendicular to $v_{n+2}$. The only unconstrained degree of freedom is the angle between the vectors $v_{n}$ and $v_{n+1}$. Thus, the vectors of the basis obey the following relations:
\begin{multline}
v_{i}\cdot v_{j}=s_i\delta_{i,j}, \quad v_{n}\cdot v_{n}=t m_-^2, \quad v_{n+1}\cdot v_{n+1}=t m_+^2,\\
v_{n+2}\cdot v_{n+2}=s\Lambda^2,\qquad v_{i}\cdot v_{n}=0,\quad v_{i}\cdot v_{n+1}=0,\\
v_{i}\cdot v_{n+2}=0,\quad v_{n}\cdot v_{n+2}=0,\quad  v_{n+1}\cdot v_{n+2}=0.
\end{multline}
To deal with the inner product $v_{n+1}\cdot v_{n+2}$ we have to be more careful and take into account the setup of the problem. In other words, we should make sure that the basis indeed spans the enhanced space. To treat all cases simultaneously, we define the primary Pohlmeyer field $\varphi$ via the equation
\begin{equation}
\partial_+Y\cdot\partial_-Y=m_+ m_-  f(\varphi).
\end{equation}
In Appendix \ref{sec:AdS} we provide details on the construction of the basis in the enhanced space and how this affects the definition of the primary Pohlmeyer field. There are three different options for the function $f$, namely
\begin{equation}\label{eq:inner_product}
f(\varphi)=\begin{cases}
\cos \varphi\\
\cosh\varphi\\
\exp \varphi
\end{cases}\qquad.
\end{equation}

In order to perform the Pohlmeyer reduction, we calculate the derivatives of the vectors constituting the basis and expand them on the basis itself, i.e.
\begin{equation}\label{eq:Pohl_derivatives}
\partial_{\pm} v_\alpha=\left(A_{\pm}\right)_{\alpha\beta}v_\beta.
\end{equation}
This equation can be expressed in matrix form as
\begin{equation}\label{eq:Pohl_derivatives_matrix}
\partial_{\pm} V=A_{\pm}V,
\end{equation}
where $V$ is the matrix having the vectors of the basis as its rows, i.e. its matrix elements are $V_{\alpha\beta}=(v_\alpha)_\beta$, see also equation \eqref{eq:V_matrix_struc}. Using the definitions of $v_{n}$ and $v_{n+1}$, along with the equations of motion \eqref{eq:EOM_NLSM}, it is straightforward to calculate that
\begin{multline}
\partial_+ v_{n+2}=v_{n+1},\quad \partial_- v_{n+2}=v_{n},\\
\partial_+ v_{n}=\partial_- v_{n+1}=-s\frac{m_+m_-}{\Lambda^2}f(\varphi)v_{n+2}.
\end{multline}
The other derivatives of $v_{n}$ and $v_{n+1}$ are obtained using the geometric constraint \eqref{eq:geometric_constraint}, the non-vanishing components of the stress energy tensor \eqref{eq:Virasoro} and the equations of motion \eqref{eq:EOM_NLSM}. In particular, these relations imply that
\begin{align}
Y\cdot \left(\partial_\pm^2 Y\right)&=-t m^2_\pm,\\
\left(\partial_\pm Y\right) \cdot \left(\partial_\pm^2 Y\right)&=0,\\
\left(\partial_\mp Y\right) \cdot \left(\partial_\pm^2 Y\right)&=m_+ m_-f^\prime(\varphi)\partial_\pm\varphi.
\end{align}
Thus, the expansion of the derivatives $\partial_\pm^2 Y$ on the basis is
\begin{multline}\label{eq:Y_second_der}
\partial^2_{\pm}Y=-st\frac{m^2_\pm}{\Lambda^2}Y+\frac{f(\varphi)f^\prime(\varphi)}{f^2(\varphi)-t^2}\partial_\pm\varphi\partial_\pm Y\\
-t\frac{m_\pm}{m_\mp}\frac{f^\prime(\varphi)}{f^2(\varphi)-t^2}\partial_\pm\varphi\partial_\mp Y+m_{\pm}a_i^{\pm} v_i,
\end{multline}
where $a_i^{\pm}$ are $2(n-1)$ unknown parameters. In the following, these parameters are grouped as the two $(n-1)\times1$ matrices ${\bf a}_\pm$. Similarly, we can calculate
\begin{multline}
\partial_\pm v_i=\frac{s_i\,a^\pm_i}{f^2(\varphi)-t^2}\left(\frac{t}{m_\pm}\partial_\pm Y-\frac{f(\varphi)}{m_\mp}\partial_\mp Y\right)\\+\left(\mathcal{A}_{\pm}\right)_{ij}s_jv_j,
\end{multline}
where $\mathcal{A}_{\pm}$ are $(n-1)\times(n-1)$ antisymmetric matrices, whose elements are given by $\left(\mathcal{A}_{\pm}\right)_{ij}=v_j\cdot\partial_\pm v_i$. Defining the matrix $\mathcal{S}$ as
\begin{equation}\label{eq:mathcal_S_definition}
\mathcal{S}=\textrm{diag}\left(s_1,\dots,s_{n-1}\right).
\end{equation} 
and putting everything together, the matrices $A^{\pm}_{\alpha\beta}$ are the following
\begin{align}
A_+&=\begin{pmatrix}
\mathcal{A}_+\mathcal{S} & \frac{f}{t^2-f^2}\frac{\mathcal{S}{\bf a}_+}{m_-} & \frac{t}{f^2-t^2}\frac{\mathcal{S}{\bf a}_+}{m_+} & {\bf 0} \\
{\bf 0}^T & 0 & 0 & -s\frac{m_+ m_-}{\Lambda^2}f\\
m_+{\bf a}_+^T & \frac{m_+}{m_-}\frac{tf^\prime}{t^2-f^2}\partial_+\varphi & \frac{ff^\prime}{f^2-t^2}\partial_+\varphi & -st\frac{m^2_+}{\Lambda^2}\\
{\bf 0}^T & 0 & 1 &0
\end{pmatrix},\\
A_-&=\begin{pmatrix}
\mathcal{A}_-\mathcal{S}  & \frac{t}{f^2-t^2}\frac{\mathcal{S}{\bf a}_-}{m_-} & \frac{f}{t^2-f^2}\frac{\mathcal{S}{\bf a}_-}{m_+} & {\bf 0} \\
m_-{\bf a}_-^T & \frac{ff^\prime}{f^2-t^2}\partial_-\varphi & \frac{m_-}{m_+}\frac{tf^\prime}{t^2-f^2}\partial_-\varphi & -st\frac{m^2_-}{\Lambda^2}\\
{\bf 0}^T & 0 & 0 & -s\frac{m_+ m_-}{\Lambda^2}f\\
{\bf 0}^T & 1 & 0 &0
\end{pmatrix},
\end{align}
where column matrices are denoted with bold symbols. We also dropped the argument of $f(\varphi)$. The compatibility condition $\partial_+\partial_-v_i=\partial_-\partial_+v_i$ implies that the matrices $A_\pm$ obey the zero curvature condition
\begin{equation}
\partial_- A_+ - \partial_+ A_-+\left[A_+,A_-\right]=0.
\end{equation}
Explicitly, the equations of motion of the Pohlmeyer fields $\varphi$, ${\bf a}_\pm$ and $\mathcal{A}_\pm$ read
\begin{align}
F_{+-}&=-\frac{f(\varphi)}{f^2(\varphi)-t^2}\left[{\bf a}_-{\bf a}_+^T-{\bf a}_+{\bf a}_-^T\right]\mathcal{S},\label{eq:gauge_Pohl_eq}\\
\mathcal{D}_\pm{\bf a}_\mp&=t\frac{f^\prime(\varphi)\partial_\mp\varphi}{f^2(\varphi)-t^2}{\bf a}_\pm,\label{eq:secondary_Pohl_eq}\\
\begin{split}\partial_+\partial_-\varphi&=\frac{{\bf a}_-^T\mathcal{S}{\bf a}_+}{f^\prime(\varphi)}-s\frac{m_+m_-}{\Lambda^2}\frac{f^2(\varphi)-t^2}{f^\prime(\varphi)}\\ &-\frac{\partial}{\partial\varphi}\left[\ln\left(\frac{f^\prime(\varphi)}{\sqrt{f^2(\varphi)-t^2}}\right)\right]\partial_+\varphi\partial_-\varphi,\label{eq:primary_Pohl_eq}\end{split}
\end{align}
where the covariant derivatives are defined as $\mathcal{D}_\pm=I_{n-1}\partial_\pm-\mathcal{S}\mathcal{A}_\pm$ and the field-strength as $F_{+-}=[\mathcal{D}_+,\mathcal{D}_-]$. Notice that the reduced theory depends only on the product $m_+m_-$. For our purposes, the function $f$ is given by \eqref{eq:inner_product}. Thus, regarding \eqref{eq:primary_Pohl_eq}, there are two options $t^2=1$ and either $f(\varphi)=\cosh\varphi$ or $f(\varphi)=\cos\varphi$,  or $t=0$ and $f(\varphi)=\exp \varphi$. In both cases the coefficient of $\partial_+\varphi\partial_-\varphi$ vanishes. Also notice that $\left(\mathcal{S}\mathcal{A}_\pm\right)^T=-\mathcal{S}\left(\mathcal{S}\mathcal{A}_\pm\right)\mathcal{S}$, implying that $\mathcal{S}\mathcal{A}_\pm$ are valued in the Lie algebra of an indefinite orthogonal group $G$. The Killing metric of this group is $\mathcal{S}$.

In this form, the Pohlmeyer reduced theory has the gauge redundancy
\begin{equation}\label{eq:gauge_redundancy}
{\bf a}_\pm\rightarrow O {\bf a}_\pm,\qquad \mathcal{S}\mathcal{A}_\pm\rightarrow O   \mathcal{S}\mathcal{A}_\pm O^{-1}+\left(\partial_\pm O\right) O^{-1},
\end{equation}
where $O^{-1}=\mathcal{S}O^T\mathcal{S}$, i.e. the matrix $O$ is an element of the indefinite orthogonal group $G$. So far, for sufficiently large $n$, we have introduced many more fields than the ones of the NLSM.\footnote{At this point, the Pohlmeyer reduced theory depends on the primary Pohlmeyer field -$1$ field-, on ${\bf a}_\pm$ -$2(n-1)$ fields- and on $\mathcal{A}_\pm$ -$(n-1)(n-2)$ fields-, for a total of $n(n-1)+1$ fields. The gauge redundancy implies that $(n-1)(n-2)/2$ fields are unphysical, leaving $n(n+1)/2$ degrees of freedom. The NLSM depends on $(n+1)$ fields.} Seemingly we are facing a more complicated problem than the original NLSM, but this is not the case. 

Finally, let us comment that so far by using the symbols $\sigma^\pm$ for the world-sheet coordinates, along with $\partial_\pm$ for the derivatives, we have implicitly assumed that the world-sheet is Minkowksi. Nevertheless, the above analysis is valid also for Euclidean  world-sheet. The main difference is that for Euclidean  world-sheet $\mathcal{A}_\pm$ and are complex and related by complex conjugation. This is also the case for  ${\bf a}_\pm$ . Also, the group of transformations in equation \eqref{eq:gauge_redundancy} is defined in the field of complex numbers, i.e. it is of the form $SO(p,q;\mathbb{C})$.

In the following we consider that the world-sheet is Minkwoski and reformulate the theory, so that it depends only on the physical degrees of freedom. As the light-like reduced theory is still conformal, whereas conformal invariance is explicitly broken for time-like and space-like reductions, the reformulation is different for these two cases. Thus, we have to treat these cases separately.

\subsection{Time-like \& space-like reductions}
For $t^2=1$ the Pohlmeyer reduction is very similar to the one of the sphere \cite{Pohlmeyer:1979ch,Katsinis:2022amo}. The idea is to introduce $n\times n$ matrices $\tilde{A}_\pm$, which constitute the components of a flat connection. So, we define $\tilde{A}_\pm$ as follows
\begin{align}
\tilde{A}_+&=\tilde{S}\begin{pmatrix}
\mathcal{A}_+ &\frac{f}{f^2-t^2} \frac{\mathcal{S}{\bf a}_+}{F(f)}\\
-\frac{f}{f^2-t^2} \frac{{\bf a}^T_+\mathcal{S}}{F(f)} & 0
\end{pmatrix},\\ \tilde{A}_-&=\tilde{S}\begin{pmatrix}
\mathcal{A}_- &\tilde{s}F(f){\mathcal{S}\bf a}_-\\
-\tilde{s}F(f){\bf a}^T_-\mathcal{S} & 0
\end{pmatrix},
\end{align}
where the matrix $\tilde{S}$ is defined as
\begin{equation}\label{eq:def_tilde_S}
\tilde{S}=\begin{pmatrix}
\mathcal{S} & 0\\
0 & \tilde{s}
\end{pmatrix},\qquad\tilde{s}=\pm1.
\end{equation}
Henceforth we drop the argument of $f(\varphi)$. Notice that by definition $\tilde{A}_\pm$ obey $\tilde{A}_\pm^T=-\tilde{S}\tilde{A}_\pm\tilde{S}$, thus they are valued in the Lie algebra of an indefinite orthogonal. The matrix $\tilde{S}$ is the Killing metric of this Lie group. We also define $\tilde{\bf{a}}_\pm$ and $\tilde{\bf{Z}}$ via the equations
\begin{equation}
\tilde{\bf{a}}_\pm=\begin{pmatrix}\bf{a}_\pm\\0\end{pmatrix},\quad \tilde{\bf{Z}}=\begin{pmatrix}
\bf{0}\\
1
\end{pmatrix}.
\end{equation}
It is a matter of algebra to show that $\tilde{\bf{a}}_\pm$ obey
\begin{align}
\tilde{\mathcal{D}}_+ \tilde{{\bf a}}_-&=t\frac{f^\prime\partial_-\varphi}{f^2-t^2}{\bf a}_++\frac{\tilde{s}}{F(f)}\frac{f}{f^2-t^2}\left(\tilde{\bf{a}}^T_-\tilde{S}\tilde{\bf{a}}_+\right) \tilde{\bf{Z}},\label{eq:tilde_a_1}\\
\tilde{\mathcal{D}}_- \tilde{{\bf a}}_+&=t\frac{f^\prime\partial_+\varphi}{f^2-t^2}{\bf a}_-+F(f)\left(\tilde{\bf{a}}^T_-\tilde{S}\tilde{\bf{a}}_+\right) \tilde{\bf{Z}},\label{eq:tilde_a_2}
\end{align}
where $\tilde{\mathcal{D}}_\pm=I_{n}\partial_\pm-\tilde{\mathcal{A}}_\pm$, while \eqref{eq:primary_Pohl_eq} assumes the form
\begin{equation}
\partial_+\partial_-\varphi=\frac{\tilde{{\bf a}}_-^T\tilde{\mathcal{S}{\bf a}}_+}{f^\prime}-s\frac{m_+m_-}{\Lambda^2}\frac{f^2-t^2}{f^\prime}\label{eq:primary_pohl_tilde}.
\end{equation}
One can show that the matrices $\tilde{\mathcal{A}}_\pm$ indeed obey the zero curvature condition
\begin{equation}\label{eq:flatness_tilde}
\partial_-\tilde{\mathcal{A}}_+-\partial_+\tilde{\mathcal{A}}_-+\left[\tilde{\mathcal{A}}_+,\tilde{\mathcal{A}}_-\right]=0,
\end{equation}
provided that the function $F$ is given by 
\begin{equation}\label{eq:sol_F}
F(f)=\pm\frac{1}{\sqrt{\tilde{s}t\left(1-f^2\right)}},\quad\text{or}\quad F(f)=\pm\frac{f}{\sqrt{\tilde{s}t\left(1-f^2\right)}}.
\end{equation}
For a given choice of $f$, which is defined in $\eqref{eq:inner_product}$, and $t$, the value of $\tilde{s}$ should be appropriate so that the square roots are well defined. As the $\pm$ signs can be absorbed by redefining ${\bf a}_\pm$, one can select
\begin{equation}
F(f)=\frac{1}{\sin\phi} \qquad F(f)=-\cot\phi
\end{equation}
for $f=\cos\varphi$, which requires $\tilde{s}t=1$, and
\begin{equation}
F(f)=\frac{1}{\sinh\phi} \qquad F(f)=-\coth\phi,
\end{equation}
for $f=\cosh\varphi$, which requires $\tilde{s}t=-1$. Notice that in both cases for either choices of $F$, appearing directly in $\tilde{A}_-$, the other one appears on the matrix $\tilde{A}_+$. Thus, the existence of two solutions restores the $+\leftrightarrow-$ symmetry.

Finally, let us comment that by promoting $\tilde{\bf{Z}}$ to a dynamical field, the reduced system, as described by \eqref{eq:tilde_a_1} - \eqref{eq:flatness_tilde}, has the following gauge redundancy: 
\begin{multline}
\tilde{\bf a}_\pm\rightarrow \tilde{O} \tilde{\bf a}_\pm,\qquad \tilde{\bf Z}\rightarrow \tilde{O} \tilde{\bf Z}\\
\tilde{\mathcal{S}}\tilde{\mathcal{A}}_\pm\rightarrow \tilde{O}   \tilde{\mathcal{S}}\tilde{\mathcal{A}}_\pm \tilde{O}^T+\left(\partial_\pm \tilde{O}\right) \tilde{O}^T,
\end{multline}
where $\tilde{O}$ belongs to the indefinite orthogonal Lie group having $\tilde{S}$ as Killing metric. Nevertheless, since $\tilde{\bf{Z}}$ in not dynamical, we can interpret the theory as a gauge fixed version of a theory originally having this symmetry. This is expected by the Lagrangian formulation of SSSGs as gauge fixed WZW models defined on an appropriate coset, which are perturbed by an integrable potential \cite{Bakas:1995bm,Miramontes:2008wt}. The group of the aforementioned symmetry coincides with the subgroup used to define the coset of the Lagrangian formulation.

In order to continue the derivation, we use equation \eqref{eq:flatness_tilde} to solve for $\tilde{\bf a}_\pm$. More specifically, if follows that $\tilde{A}_\pm=\left(\partial_\pm\tilde{O}\right)\tilde{S}\tilde{O}^T\tilde{S}$, which implies that $\tilde{\bf a}_\pm$ are given by
\begin{equation}
\tilde{\bf{a}}_+=F(f)\frac{1-f^2}{f}\mathcal{O}\partial_+{\bf Z},\qquad \tilde{\bf{a}}_-=-\frac{\tilde{s}}{F(f)}\mathcal{O}\partial_-{\bf Z},
\end{equation}
where ${\bf Z}$ is defined as
\begin{equation}
{\bf Z}=\tilde{S}\mathcal{O}^T\tilde{S}\tilde{\bf{Z}}
\end{equation}
and its norm is ${\bf Z}^T\tilde{S}{\bf Z}=\tilde{s}$. The equations of the reduced theory, taking \eqref{eq:sol_F} into account, assume the form
\begin{align}
\partial_+\partial_-\varphi=&\tilde{s}\frac{f^2-1}{ff^\prime}\partial_-{\bf Z}^T\tilde{\mathcal{S}} \partial_+{\bf Z}-s\frac{m_+m_-}{\Lambda^2}\frac{f^2-1}{f^\prime},\label{eq:EOM_phi}\\
\begin{split}\partial_+\partial_-{\bf Z}=&\frac{f^\prime}{1-f^2}\left(\frac{f}{\tilde{F}^2}\partial_+\varphi\partial_-{\bf Z}+\frac{\tilde{F}^2}{f}\partial_-\varphi\partial_+{\bf Z}\right)\\&-\tilde{s}\left(\partial_-{\bf Z}^T\tilde{\mathcal{S}} \partial_+{\bf Z}\right){\bf Z},\label{eq:EOM_Z}\end{split}
\end{align}
where $\tilde{F}=1$ or $\tilde{F}=f$ and $\tilde{\mathcal{S}}$ is defined in \eqref{eq:def_tilde_S}. The two choices for $F$, which are given in \eqref{eq:sol_F}, and equivalently of $\tilde{F}$, restore the symmetry $\sigma^+\leftrightarrow\sigma^-$. In the Lagrangian formulation of the theory these correspond to vector and axial gauging of the WZW model, see \cite{Miramontes:2008wt}. We remind the reader that the choice of $f$, $\tilde{s}$ and $t$ are correlated so that $F$ is real. As the norm of ${\bf Z}$ is constraint, the reduced theory depends on $n$ fields. Thus, the reduction eliminates one field from the NLSM. Moreover, as advertised, the reduced theory depends only on the product $m_+m_-$. Finally, notice that unless the enhanced space is 3-dimensional, so that the reduced theory consists only of the primary Pohlmeyer field, the equations we derived are non-Lagrangian. Non-local field redefinitions are required to give a Lagrangian description to the reduced theory, see \cite{Grigoriev:2007bu} and references therein. This topic is also discussed in the Appendix \ref{sec:AdS}.

\subsection{Light-like reduction}
Let us proceed to the study of the light-like reduction. Our approach is motivated by \cite{Dorn:2009kq,Hoare:2012nx}. The light-like Pohlmeyer reduction is fundamentally different than the time-like and space-like ones. The basic difference is that equations \eqref{eq:secondary_Pohl_eq} are source free. This prevents us from using the approach of the previous subsection. Equations \eqref{eq:sol_F} are not defined for $t=0$. 
As $\mathcal{S}^2=I_{n-1}$ and $\mathcal{A}_\pm^T=-\mathcal{A}_\pm$ it follows that 
\begin{equation}
\partial_\pm \left({\bf a}_\mp^T \mathcal{S}{\bf a}_\mp\right)=0,
\end{equation}
thus the norms of ${\bf a}_\pm$ depend on a sole world-sheet coordinate, i.e.
\begin{equation}
{\bf a}_\pm^T\tilde{S}{\bf a}_\pm=f_\pm\left(\sigma^\pm\right).
\end{equation}

Implementing the gauge transformation \eqref{eq:gauge_redundancy} we can simplify the problem choosing the light-cone gauge, i.e. $\mathcal{A}_-=0$. Then, equations \eqref{eq:gauge_Pohl_eq} and \eqref{eq:secondary_Pohl_eq}  assume the form
\begin{equation}\label{eq:gauge_Pohl_eq_chiral}
\mathcal{S}\partial_- \mathcal{A}_+=-\exp\left(-\varphi\right)\left[{\bf a}_-{\bf a}_+^T-{\bf a}_+{\bf a}_-^T\right]\mathcal{S}
\end{equation}
and 
\begin{equation}\label{eq:secondary_Pohl_eq_chiral}
\partial_+ {\bf a}_-=\mathcal{S}\mathcal{A}_+{\bf a}_-,\qquad \partial_- {\bf a}_+=0.
\end{equation}
These equations have the residual chiral gauge redundancy
\begin{equation}\label{eq:gauge_redundancy_chiral}
{\bf a}_\pm\rightarrow O_+ {\bf a}_\pm,\qquad \mathcal{S}\mathcal{A}_\pm\rightarrow O_+   \mathcal{S}\mathcal{A}_\pm O_+^{-1}+\left(\partial_\pm O_+\right) O_+^{-1},
\end{equation}
where $O_+=O_+(\sigma^+)$ and $O_+^{-1}=\mathcal{S}O_+\mathcal{S}$, implying $O_+\in G$. As ${\bf a}_+={\bf a}_+(\sigma_+)$, we can always use an appropriate transformation in order to set $\left({\bf a}_+\right)_i=u_+(\sigma^+)\delta_{i,(n-1)}$ if its norm is positive or negative. In the degenerate case of ${\bf a}_+$ being null, we can set $\left({\bf a}_+\right)_i=u_+(\sigma^+)\left(\delta_{i,1}+\delta_{i,(n-1)}\right)$, where $s_1s_{n-1}=-1$, see \eqref{eq:mathcal_S_definition}. From now on we assume that the norm is non-vanishing, thus we define
\begin{equation}\label{eq:a_p_decomposition}
{\bf a}_+=\begin{pmatrix}
{\bf 0} \\ u_+
\end{pmatrix},\qquad u_+=u_+(\sigma^+),
\end{equation}
where ${\bf 0}$ is a $(n-2)\times 1$ column matrix consisting of zeros. For notational convenience, we also drop the argument of $u_+(\sigma^+)$. As equation \eqref{eq:a_p_decomposition} breaks the symmetry group G to its maximal subgroup, we decompose ${\bf a}_-$ and $\mathcal{S}$, which is defined in equation \eqref{eq:mathcal_S_definition}, in a similar manner as
\begin{equation}
{\bf a}_-=\begin{pmatrix}
{\bf v} \\ v
\end{pmatrix},\qquad \mathcal{S}=\begin{pmatrix}
\bar{\mathcal{S}} & {\bf 0}\\
{\bf 0}^T & s_{n-1}
\end{pmatrix},
\end{equation}
where ${\bf v}$ is a $(n-2)\times 1$ column matrix and $\bar{\mathcal{S}}$ is a $(n-2)\times (n-2)$ matrix. Then, equation \eqref{eq:gauge_Pohl_eq_chiral} implies that $\mathcal{A}_+$ has the structure
\begin{equation}
\mathcal{A}_+=\begin{pmatrix}
{\bf 0}_{n-2} & {\bf \Lambda}\\
-{\bf \Lambda}^T & 0
\end{pmatrix},
\end{equation}
where ${\bf 0}_{n-2}$ is a $(n-2)\times (n-2)$ matrix consisting of zeros. Moreover, it follows that ${\bf \Lambda}$ satisfies the equation
\begin{equation}
\partial_-{\bf \Lambda}=-\exp\left(-\varphi\right)s_{n-1}u_+\bar{\mathcal{S}}{\bf v}.
\end{equation}
Finally, equation \eqref{eq:secondary_Pohl_eq_chiral} implies
\begin{equation}
{\bf \Lambda}=\frac{1}{v}\bar{\mathcal{S}}\partial_+{\bf v},\qquad \partial_+v=-s_{n-1}{\bf \Lambda}^T{\bf v}.
\end{equation}
The second equation is equivalent to the fact that ${\bf a}_-^T\tilde{S}{\bf a}_-=f_-\left(\sigma^-\right)$. Putting everything together, the equations of the reduced theory read
\begin{align}
\partial_-\left(\frac{1}{v}\partial_+{\bf v}\right)&=-u_+u_-\exp\left(-\varphi\right){\bf v},\label{eq:eom_light_1}\\
\partial_+\partial_-\varphi&=u_+u_-v\exp\left(-\varphi\right)-s\frac{m_+m_-}{\Lambda^2}\exp\left(\varphi\right),\label{eq:eom_light_2}
\end{align}
where we discarded the sign $s_{n-1}$ (it can be absorbed into the definition of $v$) and rescaled ${\bf v}$ and $v$ with $u_-(\sigma^-)$ so that 
\begin{equation}
{\bf a}_-= u_-\begin{pmatrix}
{\bf v} \\ v
\end{pmatrix},\qquad u_-=u_-(\sigma^-),
\end{equation}
where ${\bf v}^T\bar{\mathcal{S}}{\bf v}+s_{n-1}v^2=\bar{s}$ and $\bar{s}=\pm1$. Notice that for $t=0$ equation \eqref{eq:Y_second_der} implies that
\begin{equation}
\left(\partial^2_\pm Y\right)\cdot\left(\partial^2_\pm Y\right)=m^2_\pm {\bf a}_\pm^T\tilde{S}{\bf a}_\pm,
\end{equation}
thus
\begin{align}
\left(\partial^2_+ Y\right)\cdot\left(\partial^2_+ Y\right)&=s_{n-1}m^2_+u_+^2,\\
\left(\partial^2_- Y\right)\cdot\left(\partial^2_- Y\right)&=\bar{s}m^2_-u_-^2.
\end{align}
Using a diffeomorphism of the world-sheet coordinates, which acts as $\sigma^\pm\rightarrow f_\pm(\sigma^\pm)$ so that $f^{\prime 2}u^4_\pm=1$, we set $\left(\partial^2_\pm Y\right)\cdot\left(\partial^2_\pm Y\right)$ to constants.\footnote{Notice that no cross-terms are introduced as $\left(\partial_\pm Y\right)\cdot\left(\partial_\pm Y\right)=0$ and $\left(\partial_\pm Y\right)\cdot\left(\partial_\pm^2 Y\right)=0$.} Finally, equations \eqref{eq:eom_light_1} and \eqref{eq:eom_light_2} assume the form
\begin{align}
\partial_-\left(\frac{1}{v}\partial_+{\bf v}\right)&=\mu\exp\left(-\varphi\right){\bf v},\label{eq:eom_light_1_final}\\
\partial_+\partial_-\varphi&=\mu\left[v\exp\left(-\varphi\right)-s_{\mu}\exp\left(\varphi\right)\right],\label{eq:eom_light_2_final}
\end{align}
where we set $\varphi\rightarrow\varphi+\ln\frac{\sqrt{\vert u_+\vert\vert u_-\vert}}{\mu}$, $\mu=\left\vert\frac{m_+m_-}{\Lambda^2}\right\vert^{1/2}$ and $s_{\mu}=s\frac{m_+m_-}{\mu^2\Lambda^2}$. We also discarded the factors $\frac{u_+u_-}{\vert u_+\vert\vert u_-\vert}$ since the theory is invariant under $v\rightarrow-v$.

\section{The solution of the auxiliary system} \label{sec:Auxiliary}

As we presented the Pohlmeyer reduction of the rank 1 cosets related to the indefinite orthogonal group, we proceed to the the solution of the auxiliary system. Initially we discuss the mapping connecting  the vectors of enhanced space with the elements of coset. Then, we derive the formal solution of the auxiliary system.
\subsection{The mapping}
In order to assign an element of the coset to each vector of the enhanced space we use the mapping
\begin{equation}\label{eq:g_properties}
g=\theta\left(I_{n+2}-2s \frac{Y Y^T\eta}{\Lambda^2}\right),\quad \theta=\left(I_{n+2}-2s \frac{Y_0 Y_0^T\eta}{\Lambda^2}\right),
\end{equation}
where $I_{n}$ denotes the $n\times n$ identity matrix and $\eta$ is the metric of the enhanced space, see \eqref{eq:metric_enhanced}. Recall that the parameter $s$ is defined via the equation $Y^T \eta Y=s \Lambda^2$ and gets the values $\pm1$. Finally, $Y_0$ is a constant vector of the enhanced space, which implies that it also obeys $Y_0^T \eta Y_0=s\Lambda^2$.  It is straightforward to show that $\theta$ obeys $\theta^2=I_{n+2}$, as well as $\theta^T \eta\theta=\eta$. Similarly, $g$ has the following properties:
\begin{equation}
\bar{g}=g,\qquad g^T \eta g=\eta,\qquad g\theta g\theta=I_{n+2}.
\end{equation}
Thus $g$ is an element of the desired coset.
\subsection{The formal solution of the auxiliary system}
\label{eq:sec_solution}
Having set the ground, in this section we derive the solution of the auxiliary system \eqref{eq:auxiliary_system_lax}. Initially, we define $\Psi=g\hat{\Psi}$, so that the auxiliary system assumes the form
\begin{equation}\label{eq:auxiliary_system_hat}
\partial_\pm\hat{\Psi}=\pm\frac{2\lambda}{1\pm\lambda}\hat{j}_\pm\hat{\Psi},
\end{equation}
where the currents $\hat{j}_\pm$ are defined as $\hat{j}_\pm=-\frac{1}{2}g^{-1}\partial_\pm g$. Their explicit form is
\begin{equation}
\hat{j}_\pm=\frac{s}{\Lambda^2}\left[\left(\partial_\pm Y\right)Y^T-Y\partial_\pm Y^T\right]\eta.
\end{equation}
Interestingly enough, equation \eqref{eq:auxiliary_system_hat} implies that $\hat{\Psi}(0)$ is constant. This is a crucial remark in order to gain intuition about the form of the structure of the solution of the auxiliary system. The only linear system of equation that is somehow related to the problem is \eqref{eq:Pohl_derivatives_matrix}. It is natural to guess that the solution depends linearly on $V$, which is $\lambda$ independent, and the rest of the solution reduced to $V^{-1}$ for $\lambda=0$. This is close, but it is not correct. It turns out the correct ansatz is
\begin{equation}\label{eq:Psi_tilde_def}
\hat{\Psi}=V^{-1}\Delta^{-1}\tilde{\Psi},
\end{equation}
where the matrix $\Delta$ is defined as
\begin{equation}
\Delta=\begin{pmatrix}
I_{n-1} & 0 & 0 & 0\\
0 & \frac{1+\lambda}{1-\lambda}& 0 &0 \\
0 & 0 & \frac{1-\lambda}{1+\lambda}& 0\\
0 & 0 & 0 &1
\end{pmatrix},
\end{equation}
while the matrix $V$ is constructed by the vectors of the basis used in the Pohlmeyer reduction, see \eqref{eq:Pohl_derivatives_matrix}. Taking equations \eqref{eq:Pohl_derivatives} into account, the auxiliary system assumes the form
\begin{equation}
\partial_\pm\tilde{\Psi}=\Delta\left[A^\pm\pm\frac{2\lambda}{1\pm\lambda}V\hat{j}_\pm V^{-1}\right]\Delta^{-1}\tilde{\Psi}.
\end{equation}
It is a matter of algebra to show that the above pair of equations assumes the form
\begin{equation}\label{eq:Psi_tilde_final}
\partial_\pm\tilde{\Psi}=\left[A^\pm\vert_{m_\pm\rightarrow\frac{1\mp\lambda}{1\pm\lambda}m_\pm}\right]\tilde{\Psi}.
\end{equation}
For completeness we provide some details of this calculation in Appendix \ref{sec:details}. Comparing equations \eqref{eq:Psi_tilde_final} and \eqref{eq:Pohl_derivatives_matrix} it follows that
\begin{equation}\label{eq:Psi_tilde_solution}
\tilde{\Psi}=V\vert_{m_\pm\rightarrow\frac{1\mp\lambda}{1\pm\lambda}m_\pm}.
\end{equation}
Putting everything together, the solution of the auxiliary system reads
\begin{equation}\label{eq:Psi_solution}
\Psi(\lambda)=gV^{-1}\Delta^{-1}\left(V\vert_{m_\pm\rightarrow\frac{1\mp\lambda}{1\pm\lambda}m_\pm}\right)
\end{equation}
and coincides with the result of \cite{Katsinis:2022amo}, which is expected as the $O(n)$ NLSM is a special case of the models under consideration. Let us also mention that the structure of the solution was actually guessed using the result of the $O(3)$ NLSM  \cite{Katsinis:2020avk}, which was expressed in terms of the embedding functions.
\subsection{Properties of the solution}
Having solved the auxiliary system, we present some basic properties of the solution. To begin with, it is straightforward to show that the solution is normalized so that
\begin{equation}
\Psi(0)=g.
\end{equation}
Of course, the set of solutions of the auxiliary system is of the form $\Psi(\lambda)C(\lambda)$ where $C(\lambda)$ is a constant matrix, which may depend on $\lambda$.

As $g$, belonging to a coset, obeys the relations in \eqref{eq:g_properties}, it turns out that there are some consistency conditions for the solution of the auxiliary system \eqref{eq:auxiliary_system_lax}. Since $g$ is real it follows that for Minkowski world-sheet $\Psi(\lambda)$ has to satisfy
\begin{equation}
\bar{\Psi}(\bar{\lambda})=\Psi(\lambda).
\end{equation}
Given that the NLSM solution $Y$ is real function of the real parameters $m_\pm$, the solution \eqref{eq:Psi_solution} satisfies this constraint. In the case of Euclidean world-sheet, since $\partial_\pm$ are interchanged by complex conjugation, the analogous equation is
\begin{equation}
\bar{\Psi}(\bar{\lambda})=\Psi(-\lambda).
\end{equation}
The NLSM solution $Y$ is function of the complex parameters $m_\pm$. Notice that $m_\pm$ are related by complex conjugation, thus the constraint is indeed satisfied. 

Since  $g^T \eta g=\eta$ it follows that
\begin{equation}
\Psi\eta\Psi^T =\Psi^T \eta\Psi=\eta.
\end{equation}
Using equations \eqref{eq:V_matrix_struc} and \eqref{eq:V_inv_matrix_struc}, along with equations \eqref{eq:inverse_mat_1}-\eqref{eq:inverse_mat_3} and the properties of the vectors $v_\alpha$, which constitute the basis, it is straightforward to show that
\begin{multline}
\Delta^{-1}\left(V\vert_{m_\pm\rightarrow\frac{1\mp\lambda}{1\pm\lambda}m_\pm}\right)\eta\left(V^T\vert_{m_\pm\rightarrow\frac{1\mp\lambda}{1\pm\lambda}m_\pm}\right)\\
\cdot\Delta^{-1}\left(V^{-1}\right)^T=V\eta,
\end{multline}
which proves that $\Psi\eta\Psi^T =\eta$. One can prove the other equality in a similar manner.

Finally, as $\theta g\theta g=I_{n+2}$ it follows that $g\theta\Psi(1/\lambda)\theta$ belongs to the set of solutions of the auxiliary system. Thus, one has to impose the constraint
\begin{equation}
g\theta\Psi(1/\lambda)\theta=\Psi(\lambda)M(\lambda),
\end{equation}
where $M(\lambda)$ is a constant, in general $\lambda$-dependent, matrix. Using equation \eqref{eq:Psi_tilde_def} along with $\Psi=g\hat{\Psi}$ we obtain
\begin{equation}
\mathcal{I}\tilde{\Psi}(1/\lambda)\theta=\tilde{\Psi}(\lambda)M(\lambda),\quad \mathcal{I}=\begin{pmatrix}
I_{n-1} & 0 \\
0 & -I_3
\end{pmatrix}.
\end{equation}
Equation \eqref{eq:Psi_tilde_solution} implies that the above constraint is equivalent to the fact that the following equation
\begin{equation}
\mathcal{I}V\vert_{m_\pm\rightarrow-m_\pm}=V M(\lambda),
\end{equation}
is true for any $m_{\pm}$ for some constant matrix $M$.
\section{On the conserved charges}\label{sec:charges}
Having obtained the formal solution of the auxiliary system, let us calculate the eigenvalues of the monodromy matrix. Equation \eqref{eq:monodromy_matrix} implies that the  monodromy matrix reads
\begin{multline}
\label{eq:monodromy_sol}
T(\sigma_f,\sigma_i;\lambda)=g(\tau,\sigma_f)V^{-1}(\tau,\sigma_f)\Delta^{-1}(\lambda)\\
\cdot\left(V(\tau,\sigma_f)\vert_{m_\pm\rightarrow\frac{1\mp\lambda}{1\pm\lambda}m_\pm}\right)\left(V^{-1}(\tau,\sigma_i)\vert_{m_\pm\rightarrow\frac{1\mp\lambda}{1\pm\lambda}m_\pm}\right)\\
\cdot\Delta(\lambda)V(\tau,\sigma_i)g^{-1}(\tau,\sigma_i).
\end{multline}
Recall that equation \eqref{eq:monodromy_matrix}, and consequently this equation, is valid for either periodic boundary conditions i.e. $Y(\tau,\sigma_i)=Y(\tau,\sigma_f)$, or for $\mathcal{L}_\tau(\tau,\sigma_i)=\mathcal{L}_\tau(\tau,\sigma_f)=0$, implying $g(\tau,\sigma_i)$ and $g(\tau,\sigma_f)$ are constants and equivalently $Y(\tau,\sigma)$ obeys Dirichlet boundary conditions.

Notice that $\tau$ is the world-sheet temporal coordinate, which for a general manifold is non-trivially connected to the target space temporal coordinate $t$. Considering the $O(n)$ model, i.e. stings on $\mathbb{R}_t\times\textrm{S}^{n-1}$, the target space time $t$ is given by $t=m_+\sigma^++m_-\sigma^-$. In this case one can boost the solution in order to achieve $t\sim \tau$. 

Leaving this subtlety aside and assuming periodic boundary conditions, the eigenvalues of the monodromy matrix coincide with the eigenvalues of the matrix
\begin{equation}\label{eq:monodromy_eigs}
\mathcal{T}=\left(V(\tau,\sigma_f)V^{-1}(\tau,\sigma_i)\right)\vert_{m_\pm\rightarrow\frac{1\mp\lambda}{1\pm\lambda}m_\pm}.
\end{equation}
Notice that the solution is periodic for some specific $m_+$ and $m_-$, thus $g(\tau,\sigma_f)=g(\tau,\sigma_i)$ and $V(\tau,\sigma_f)=V(\tau,\sigma_i)$. The rescaling \eqref{eq:m_rescalling} spoils the periodic boundary conditions. This is the origin of the non-trivial monodromy matrix. Notice that in equation \eqref{eq:monodromy_eigs}, one should first rescale $m_\pm$ and then substitute the value of $\sigma$.
\subsection{The example of elliptic strings in $\mathbb{R}_t\times\textrm{S}^2$.}
As an indicative example let us present the case of $O(3)$ NLSM. For this model the matrix $V$ is given by $V=\begin{pmatrix}
\partial_- \bf{X} & \partial_+ \bf{X} & \bf{X}
\end{pmatrix}$ or explicitly
\begin{multline}\label{eq:V_decomposition}
V=\begin{pmatrix}
\cos\phi & -\sin\phi & 0\\
\sin\phi & \cos\phi & 0\\
0 & 0 & 1
\end{pmatrix}\begin{pmatrix}
\cos\theta & 0 & \sin\theta\\
0 & 1 & 0\\
-\sin\theta & 0 & \cos\theta
\end{pmatrix}\\
\cdot\begin{pmatrix}
\partial_-\theta & \partial_+\theta & 0\\
\sin\theta\partial_-\phi & \sin\theta\partial_+\phi & 0\\
0 & 0 & 1
\end{pmatrix}.
\end{multline}

Considering the case of elliptic strings in $\mathbb{R}_t\times\textrm{S}^2$, constructed in \cite{Katsinis:2018zxi} \footnote{In order to follow the conventions of this work , we use the coordinates $\xi^\pm=\frac{1}{2}\left(\xi^1\pm\xi^0\right)$ implying $\partial_\pm=\partial_1\pm\partial_0$. This section is self-contained, but the reader may concult \cite{Katsinis:2018zxi} for additional details.}, the solution reads
\begin{align}
t&=m_+\xi^++m_-\xi^-\label{eq:elliptic_time}\\
\sin\theta&=\sqrt{\frac{\wp(\xi^1+i\omega_2)-\wp(a)}{x_1-\wp(a)}}\\
\cos\theta&=\sqrt{\frac{x_1-\wp(\xi^1+i\omega_2)}{x_1-\wp(a)}}\\
\phi&=\ell\xi^0-\Phi(\xi^1;a)\label{eq:elliptic_phi},
\end{align}
where $\ell=\sqrt{x_1-\wp(a)}$ and $\wp(z;g_2,g_3)$ is the Weierstrass elliptic function, $\omega_2$ is the imaginary half-period and $\Phi(\xi^1;a)$ is the Bloch phase of the L\'ame eigenfunctions, which obeys
\begin{equation}\label{eq:Lame_der}
\Phi^\prime(\xi^1;a)=\frac{i}{2}\frac{\wp^\prime(a)}{\wp(\xi^1+i\omega_2)-\wp(a)}.
\end{equation}
The periodic properties of $\wp(\xi)$ are inherited from the Pohlmeyer reduced theory, thus they remain invariant under the rescaling $\eqref{eq:m_rescalling}$. Specifically, the invariants $g_2$ and $g_3$ read
\begin{equation}\label{eq:weier_invariants}
g_2=\frac{E^2}{3}+m_+^2m_-^2,\quad g_3=\frac{E}{3}\left(\left(\frac{E}{3}\right)^2-m_+^2m_-^2\right),
\end{equation}
where $ E\geq m_+m_-$ and $m_+m_-<0$. For completeness let us also mention that the roots of the cubic polynomial $Q(x)=4x^3-g_2x-g_3$ are
\begin{equation}
x_1=\frac{E}{3},\quad x_2=-\frac{E}{6}-\frac{m_+m_-}{2},\quad x_3=-\frac{E}{6}+\frac{m_+m_-}{2}.
\end{equation}
The parameter $a$ is defined via the equations
\begin{align}
\wp(a)&=-\frac{E}{6}-\frac{m_+^2+m_-^2}{4},\\
\frac{\wp^\prime(a)}{\ell}&=i\frac{m_+^2-m_-^2}{2}.
\end{align}
In the terminology of \cite{Katsinis:2018zxi}, the above solution is characterized as static. Translationally invariant solutions are obtained by applying the $\xi^0\leftrightarrow\xi^1$ duality at the coordinates $\theta$ and $\phi$. Without loss of generality we will analyze only static solutions. The treatment of translationally invariant ones is similar. For the purposes of this works it is sufficient to know that $\wp(\xi)$ and $\wp^\prime(\xi)$ are periodic under $\xi\rightarrow\xi+2\omega_1$, where $\omega_1$ is the real half-period. Notice that $\Phi(\xi;a)$ is quasi-periodic (but its derivative is periodic; see equation \eqref{eq:Lame_der}) satisfying
\begin{equation}\label{eq:Lame_quasi}
\delta\Phi=\Phi(\xi+2\omega_1;a)-\Phi(\xi;a)=2i\left(\zeta\left(a\right)\omega_1-\zeta\left(\omega_1\right)a\right).
\end{equation}
In order to introduce the target space time coordinate we substitute
\begin{equation}\label{eq:boost}
\xi^0=\gamma\left(\tau-\beta\sigma\right)\qquad \xi^1=\gamma\left(\sigma-\beta\tau\right)
\end{equation}
where
\begin{equation}
\beta=\frac{m_++m_-}{m_+-m_-},\quad \gamma=\frac{\vert m_+-m_-\vert}{2\sqrt{-m_+m_-}}.
\end{equation}
For $m_+>m_-$ it follows that $t=\sqrt{-m_+m_-}\tau$. In the boosted coordinates, the solution is periodic for $\sigma\rightarrow 
\sigma+\frac{2\omega_1}{\gamma}$ for $E>-m_+m_-$ (rotating solutions) and $\sigma\rightarrow \sigma+\frac{4\omega_1}{\gamma}$ for $-m_+m_->E>m_+m_-$ (oscillating solutions). In order to satisfy periodic boundary conditions it is required that $\sigma_f-\sigma_i=\frac{2n\omega_1}{\gamma}$, where $n\in\mathbb{N}$ (notice that $n$ is even for oscillating solutions), while $a$ gets discrete values so that
\begin{equation}
\delta\phi=n\delta\phi_1(a)=2m\pi,\qquad m\in\mathbb{N},
\end{equation}
where
\begin{equation}\label{eq:delta_phi_def}
\delta\phi_1(a)=2i\left[\zeta\left(\omega_1\right)a-\zeta\left(a\right)\omega_1\right]-2\omega_1\sqrt{x_1-\wp(a)}\beta.
\end{equation}
This is the angular opening of the elliptic strings, see section 6 of \cite{Katsinis:2018zxi}. In order to derive the above equation, one uses the quasi-periodicity property \eqref{eq:Lame_quasi}.

In order to calculate $\mathcal{T}$ we first rescale $m_\pm$ and then we boost the solution to the static gauge. Rescaling $m_\pm$ amounts to $a\rightarrow\tilde{a}$, where $\tilde{a}$ is defined via the equations
\begin{align}
\wp(\tilde{a})&=-\frac{E}{6}-\frac{m_+^2}{4}\left(\frac{1-\lambda}{1+\lambda}\right)^2-\frac{m_-^2}{4}\left(\frac{1+\lambda}{1-\lambda}\right)^2,\label{eq:wpa_tilde}\\
\frac{\wp^\prime(\tilde{a})}{\sqrt{\Delta}}&=i\left(\frac{m_+^2}{2}\left(\frac{1-\lambda}{1+\lambda}\right)^2-\frac{m_-^2}{2}\left(\frac{1+\lambda}{1-\lambda}\right)^2\right).\label{eq:wppa_tilde}
\end{align}
The parameter $\Delta=x_1-\wp(\tilde{a})$ generalizes $\ell^2$, but it is no longer positive definite. As $\theta$ and $\partial_\pm\phi$ are periodic even for the rescaled $m_\pm$, using \eqref{eq:V_decomposition} we obtain
\begin{equation}\label{eq:mathcalT_res}
\mathcal{T}=\begin{pmatrix}
\cos\delta\phi & -\sin\delta\phi & 0\\
\sin\delta\phi & \cos\delta\phi & 0\\
0 & 0 & 1
\end{pmatrix},
\end{equation}
where $\delta\phi=\phi(\sigma_f)-\phi(\sigma_i)$. Taking \eqref{eq:delta_phi_def} into account, it turns out that
\begin{equation}
\delta\phi=2n\delta\phi_1(\tilde{a}).
\end{equation}
Thus, $\delta\phi$ is the analytic continuation of the angular opening of the elliptic strings. Notice that even though $\beta$ is a function of $a$, namely
\begin{equation}
\beta=\sqrt{\frac{x_3-\wp(a)}{x_2-\wp(a)}},
\end{equation}
it is not to be analytically continuated. The origin of this term is the boost \eqref{eq:boost}, which concerns the original NLSM solution. Finally, the eigenvalues of $\mathcal{T}$ are $1$ and $e^{\pm i\delta\phi}$.

In order to verify the above calculation we can use the results of \cite{Katsinis:2018ewd}. In this work the solution of the auxiliary system corresponding to the elliptic strings in $\mathbb{R}_t\times\textrm{S}^2$ was obtained explicitly. Keeping the presentation to the absolutely necessary, the solution of the auxiliary system is
\begin{equation}
\Psi=\theta U\theta \tilde{E}\begin{pmatrix}
\cos\tilde{\phi}& \sin\tilde{\phi} & 0\\
\sin\tilde{\phi} & -\cos\tilde{\phi} & 0\\
0& 0 & 1
\end{pmatrix},
\end{equation}
where the matrix $\tilde{E}$ is defined as
\begin{equation}
\tilde{E}=\begin{pmatrix}
{\bf e_1} & {\bf e_2}  & {\bf e_3}
\end{pmatrix},
\end{equation}
the columns ${\bf e_i}$ constitute an orthonormal basis, $\theta$ is a constant matrix defined as $\theta=\textrm{diag}(1,1,-1)$, the angle $\tilde{\phi}$ is $\tilde{\phi}=\sqrt{\Delta}\xi^0-\Phi(\xi^1;\tilde{a})$ and $U=U_2U_1$. The matrices $U_1$ and $U_2$ are defined as
\begin{equation}
U_1=\begin{pmatrix}
\cos\theta& 0 & \sin\theta\\
0 & 1 & 0\\
-\sin\theta& 0 & \cos\theta
\end{pmatrix},\thickspace U_2=\begin{pmatrix}
\cos\phi& -\sin\phi & 0\\
\sin\phi & \cos\phi & 0\\
0& 0 & 1
\end{pmatrix},
\end{equation}
where the angles $\theta$ and $\phi$, along with the target space time, are defined in equations \eqref{eq:elliptic_time}-\eqref{eq:elliptic_phi}. The parameter $\tilde{a}$ is defined via \eqref{eq:wpa_tilde} and \eqref{eq:wppa_tilde}, and $\Delta=x_1-\wp(\tilde{a})$. In addition, the elements of the matrix $\tilde{E}$ depend only on $\xi^1$ and are periodic under $\xi^1\rightarrow\xi^1+2\omega_1$, where $\omega_1$ is the real half-period corresponding to the invariants \eqref{eq:weier_invariants}.

Boosting the solution to the static gauge via \eqref{eq:boost} and using equation \eqref{eq:monodromy_matrix} we obtain
\begin{equation}
T=\theta U(\sigma_f)\theta\tilde{E}(\sigma_f)\mathcal{T}\tilde{E}^{-1}(\sigma_i)\theta U^{-1}(\sigma_i)\theta,
\end{equation}
where $\mathcal{T}$ is given by \eqref{eq:mathcalT_res}. As both matrices $U$ and $\tilde{E}$ are periodic, it is evident that the eigenvalues of $T$ coincide with the eigenvalues of $\mathcal{T}$. Thus, as expected the conserved charges calculated using the methods of this work coincide with the ones calculated using the explicit results of \cite{Katsinis:2018ewd}.
\section{Discussion}

In this work we demonstrated that all rank 1 symmetric spaces related to indefinite orthogonal groups have a property that has been unnoticed for more than 4 decades. One can obtain the formal solution of the auxiliary system, which imprints the integrability of the models, corresponding to an arbitrary NLSM solution. The solution of the auxiliary system is closely related to the Pohlmeyer reduction, revealing another link between the NLSM and its reduced theory. In view of this result the dressing method is essentially the implementation of a non-linear superposition principle. The seed solution is combined appropriately with a ``virtual'' one in order to obtain a new solution. The ``virtual'' solution is constructed systematically from the seed one by the substitution \eqref{eq:m_rescalling}. This implies that the ``virtual'' solution solves the equations of motion, shares the same Pohlmeyer counterpart with the seed solution, but belongs to the complexification of the coset. Even though we focus on theories having a Minkowski world-sheet, our results apply for Euclidean world-sheet too.

Since the relation between the NLSM and the reduced theory, along with their relevance for AdS/CFT correspondence, is analyzed extensively in the Introduction, in the following we mostly present some future directions and ideas, which to our opinion are worth exploring.

First of all, it would be interesting to study more extensively the Pohlmeyer reduced. The study of the B\"acklund transformations of the reduced theory following the approach of \cite{Katsinis:2020avk} is of particular interest. More specifically, one could show explicitly that the Pohlmeyer reduced theory corresponding to the dressed solution, which is constructed using the simplest dressing factor, is related with a B\"acklund transformation to the Pohlmeyer reduced theory of the seed solution. As a byproduct, the relation between the location of the poles of the dressing factor will be related to the parameter of the B\"acklund transformation. Finally, a more extended study of the reduced theories in the case of Euclidean world-sheet would be compelling.

Our results introduce a new perspective on long strings, i.e. strings in the decompactification limit of their world-sheet. Considering them as dressed versions of short strings, this work implies that the dressed strings are superpositions of the short strings. Thus, their properties are inherited from the short strings. This point of view is closely related to the study of properties of dressed strings along the lines of \cite{Katsinis:2019oox,Katsinis:2019sdo}. These works relate the stability of classical strings to the dressed version of these strings. It turns out that the configurations are unstable whenever superluminal solitons can propagate on the background of the Pohlmeyer counterpart of them. Nevertheless, one should notice that the boundary conditions are crucial for such investigations; therefore, an abstract study may be very challenging. On the contrary our results make the stability analysis of specific string solutions pretty straightforward.

As this work indicates that the non-linear superposition concerns a wide class of symmetric spaces, its natural to wonder whether this could be the case for even more theories. Before considering a generic symmetric space, a prime candidate for such a study would be the $\mathbb{CP}^n$ NLSM. In view of recent developments regarding the construction of integrable 2-dimensional theories, the study of $\lambda$ \cite{Sfetsos:2013wia}, Yang-Baxter / $\eta$ \cite{Klimcik:2008eq,Delduc:2013fga,Delduc:2013qra}, bi-Yang-Baxter \cite{Klimcik:2014bta} deformations and generalizations of them is definitely is interesting. Regarding the $\lambda$-deformations, the Pohlmeyer reductions has been presented in \cite{Hollowood:2014rla}, providing a starting point for the aforementioned study.

The original and the reduced theory are closely related at the classical level. It is natural to wonder what happens when quantum corrections are taken into account. In the case of $O(3)$  NLSM and the sine-Gordon equations, the quantum theories are very different. As the conformal symmetry of the NLSM is anomalous, the discrepancy is not surprising. Turning this argument on its head, it could be the case that a theory, which is conformal at the quantum level, such as superstrings on AdS$_5\times$S$^5$ is somehow related to the quantized reduced theory. The natural regime for such an investigation is the BMN particle \cite{Berenstein:2002jq}, which is related to the vacuum of the reduced theory. One can show that the S-matrices of both theories are related \cite{Hoare:2009fs,Hoare:2010fb,Hoare:2011fj}. The same is true for the partition functions too \cite{Hoare:2009rq,Iwashita:2010tg,Iwashita:2011ha}.

These results motivate various studies related to this work. As supersymmetry seems to bind together the NLSM and the reduced theory, generalizing the presented construction in the case of symmetric spaces related to supercosets would certainly be interesting. First of all, supercosets have only been studied group theoretically. Even establishing such a reduction as an embedding problem is intriguing. The supersymmetric extension of the non-linear superposition could be used to further probe the spectral problem of AdS/CFT at the classical level by constructing explicitly the quasi-momenta and the spectral curve of \cite{Beisert:2005bm}.

Given a seed solution, the superposition creates an infinite tower of classical vacua of the theory. Of course, perturbation theory around the seed solution fails to describe these vacua, thus the presented construction is essential in order describe them. In the spirit of the previous paragraph one could study the semi-classical quantization of the theory around different vacua, which essentially correspond to different instanton charges. This could also shed light on the relation between the NLSM and the reduced theory in the quantum regime. A related subject concerns the study of admissible configuations of the path integral \cite{Krichever:2020tgp}. 

Finally, although we mentioned the spectral curve \cite{Beisert:2005bm} in order to motivate the introduction of supersymmetry, even without supersymmetry the spectral curve deserves a thorough study. A direct consequence of the construction we presented is that we can calculate explicitly the monodromy matrix. As a first step, it would be interesting to calculate explicitly  the quasi-momenta corresponding to a specific solution. As we analyzed in section \ref{sec:charges}, in the case of periodic boundary conditions the eigenvalues of the monodromy matrix coincide with the eigenvalues of the matrix
\begin{equation}
\mathcal{T}=\left(V(\tau,\sigma_f)V^{-1}(\tau,\sigma_i)\right)\vert_{m_\pm\rightarrow\frac{1\mp\lambda}{1\pm\lambda}m_\pm}.
\end{equation}
The point to make is that even though the NLSM solution and correspondingly the coset element $g$ and the matrix $V$ satisfy periodic boundary conditions for $m_{\pm}$, the rescaling of these parameters spoils the periodic boundary conditions. The non-trivial monodromy emerges because the NLSM solution is periodic for specific $m_{\pm}$. Of course, one should perceive the notation of \eqref{eq:monodromy_eigs} as rescaling $m_{\pm}$ first and then multiplying the two terms. In any case, the quasi-momenta are directly linked to the embedding of the NLSM solution in the enhanced space. It would be interesting to calculate them explicitly in terms of the embedding functions or relate them with the reduced theory. For instance, in the case of $O(3)$ NLSM, based on the analysis in the case of elliptic strings, it is natural to assume that the eigenvalues are $1$ and $e^{\pm i\delta\phi}$, where $\delta\phi=\phi(\sigma_f)-\phi(\sigma_i)$ and $\phi$ is the angular angle corresponding to the rescaled parameters $m_\pm$. 

Naturally, identifying in field theory the analogue of the construction we presented is very interesting. Even though it seems far-fetched, AdS/CFT correspondence motivates the existence of a similar structure as the one that was presented in the work.

\begin{acknowledgments}
The work of the author was supported by FAPESP grant 2021/01819-0.
\end{acknowledgments}

\appendix
\section{On the definition of the basis in the enhanced space}
\label{sec:AdS}
In this appendix we apply the formalism developed in section \ref{sec:Pohlmeyer_Reduction}. We restrict our analysis to setups which are relevant for AdS/CFT. First of all we have to clarify the definition of the primary Pohlmeyer field along with the construction of the basis in the enhanced space. In order to do so we have to distinguish between time-like world-sheets and space-like ones, for instance see \cite{Dorn:2009kq}, which depends on the world-sheet metric. The situation is more complicated when the space is of indefinite signature.

As mentioned the type of reduction affects the construction of the basis in the enhanced space. One should make sure that the basis indeed spans the enhanced space and that the inner product of $\partial_+Y$ and $\partial_-Y$ is appropriately defined. The basis naturally splits the enhanced space as the direct product $\mathcal{M}_3\times\mathcal{M}_{n-1}$, where $\mathcal{M}_3$ is spanned by $Y$ and $\partial_\pm Y$. In the case of AdS the enhanced space is $\mathbb{R}^{(2,n)}$, while in the case of dS it is  $\mathbb{R}^{(1,n+1)}$. Thus, in the former case the basis should consist of 2 time-like vectors and $n$ space-like, while in the latter of $1$ time-like and $n+1$ space-like ones. In the following we analyse the definition of the primary Pohlmeyer field for each type of reduction in the case of AdS. Similar analysis also applies to dS space, as well as on all other rank 1 symmetric spaces based on indefinite orthogonal groups. As a warm up we first analyze the case of sphere for both choices of world-sheet signatures.

For Minkowski world-sheets we use the following conventions
\begin{equation}
\sigma^\pm=\frac{1}{2}\left(\sigma\pm\tau\right),\quad \partial_\pm=\partial_\sigma\pm\partial_\tau,
\end{equation}
while for Euclidean ones 
\begin{equation}
\sigma^\pm=\frac{1}{2}\left(\sigma\pm i\tau\right),\quad \partial_\pm=\partial_\sigma\mp i\partial_\tau.
\end{equation}

\subsection{Sphere with Minkowski world-sheet}
We begin our presentation by discussing the basis in the case the target space is a sphere and the world-sheet is Minkowski. The inner product $\partial_+X \cdot \partial_-X$ satisfies the Cauchy-Schwartz inequality 
\begin{equation}
\partial_+X \cdot \partial_-X\leq \left\vert \partial_+X\right\vert\left\vert \partial_-X\right\vert=\vert m_+ m_-\vert.
\end{equation}
Thus, the only consistent definition of the Pohlmeyer field is
\begin{equation}
\partial_+X \cdot \partial_-X=m_+m_-\cos\varphi.
\end{equation}
Of course, the norms of all vectors spanning $\mathcal{M}_{n-1}$ are positive. It is straightforward to calculate that
\begin{align}
\partial_\sigma X\cdot\partial_\sigma X&=\frac{m_+^2+m_-^2}{4}+\frac{m_+m_-}{2}\cos\varphi,\\
\partial_\tau X\cdot\partial_\tau X&=\frac{m_+^2+m_-^2}{4}-\frac{m_+m_-}{2}\cos\varphi,\\
\partial_\sigma X\cdot\partial_\tau X&=\frac{m_+^2-m_-^2}{4}.
\end{align}
Both vectors $\partial_\sigma X$ and $\partial_\tau X$ have positive definite norm as required. 
\subsection{Sphere with Euclidean world-sheet}
Even though we will not discuss Euclidean world-sheets further, we present the basis used in the reduction when the target space is a sphere. In this case it follows that
\begin{equation}
\partial_+X\cdot \partial_-X=\partial_\sigma X\cdot\partial_\sigma X+\partial_\tau X\cdot\partial_\tau X\geq 0
\end{equation} 
We define the Pohlmeyer field as
\begin{equation}
\partial_+X \cdot \partial_-X=m_+m_-\cosh\varphi.
\end{equation}
The right-hand-side is positive definite since the parameters $m_\pm$ are related by complex conjugation, i.e. $m_\pm=m_R\pm i m_I$, where $m_{R}$ and $m_{I}$ are real. Again, it is straightforward to calculate that
\begin{align}
\partial_\sigma X\cdot\partial_\sigma X&=\frac{m_+m_-}{2}\cosh\varphi+\frac{m_+^2+m_-^2}{4}\\
&=m_R^2\cosh^2\frac{\varphi}{2}+m_I^2\sinh^2\frac{\varphi}{2},\\
\partial_\tau X\cdot\partial_\tau X&=\frac{m_+m_-}{2}\cosh\varphi-\frac{m_+^2+m_-^2}{4}\\
&=m_R^2\sinh^2\frac{\varphi}{2}+m_I^2\cosh^2\frac{\varphi}{2},\\
\partial_\sigma X\cdot\partial_\tau X&=\frac{m_-^2-m_+^2}{4i}=-m_R m_I.
\end{align}
Both vectors $\partial_\sigma X$ and $\partial_\tau X$ have positive definite norm as required. 
\subsection{AdS time-like reduction \& Minkowski world-sheet}
In this section we analyze the time-like reduction of AdS spaces. We remind the reader that this case corresponds to $t=-1$, i.e. $\partial_\pm Y\cdot \partial_\pm Y=-m^2_\pm$. Consider an arbitrary vector $X$ in the sub-manifold spanned by $\partial_\pm Y$. It will be of the form
\begin{equation}
X=a\partial_+Y+b\partial_-Y.
\end{equation}
It follows that
\begin{multline}
X\cdot X=-\left(\vert a m_+\vert-\vert b m_-\vert\right)^2\\
-2\vert a m_+\vert\vert b m_-\vert\left(1-\frac{ab\partial_+Y\cdot\partial_-Y}{\vert a m_+\vert\vert b m_-\vert}\right).
\end{multline}
If the last parenthesis is positive, all vectors in $\mathcal{M}_3$ are of negative norm, implying that for $f=\cos\varphi$ it follows that $\mathcal{M}_3=\mathbb{R}^{(3,0)}$. Thus, for $f=\cos\varphi$ the basis would not span the enhanced space $\mathbb{R}^{(2,n)}$. In this case it is necessary to use $f=\cosh\varphi$, i.e.
\begin{equation}
\partial_+Y\cdot\partial_-Y=m_+ m_-\cosh \varphi,
\end{equation}
so that $\mathcal{M}_3$ can be $\mathbb{R}^{(2,1)}$. We could have introduced another sign, such as $t$ in this definition, but we opt to take advantage of the transformation $\alpha\rightarrow \alpha+i\pi$ in order to deal with this sign. Also notice that
\begin{align}
\partial_\sigma Y\cdot\partial_\sigma Y&=-\frac{m_+^2+m_-^2}{4}+\frac{m_+m_-}{2}\cosh\varphi,\\
\partial_\tau Y\cdot\partial_\tau Y&=-\frac{m_+^2+m_-^2}{4}-\frac{m_+m_-}{2}\cosh\varphi,\\
\partial_\sigma Y\cdot\partial_\tau Y&=-\frac{m_+^2-m_-^2}{4}.
\end{align}
The consistency of the basis requires 
\begin{equation}
\cosh\varphi\geq \frac{m_+^2+m_-^2}{2\vert m_+m_-\vert},
\end{equation}
so that one of the vectors $\partial_\sigma Y$ and $\partial_\tau Y$ is time-like, while the other one is space-like

In this case, it follows that $\mathcal{S}=I_{n-1}$, i.e. the Killing metric of $SO(n-1)$. As $f=\cosh\varphi$, it is required that $\tilde{s}t=-1$. For the time-like reduction $t=-1$, thus $\tilde{s}=1$. As a result $\tilde{\mathcal{S}}=I_{n}$, which is the Killing metric of $SO(n)$. The reduced theory is formulated as the $SO(1,n)/SO(n)$ perturbed WZW model, see section 5.2 of \cite{Miramontes:2008wt}. 

For completeness, we also present the equations of the reduced theory, which read
\begin{align}
\partial_+\partial_-\varphi=&\tanh\varphi\,\partial_-{\bf Z}^T\tilde{\mathcal{S}} \partial_+{\bf Z}+\frac{m_+m_-}{\Lambda^2}\sinh\varphi,\label{AdS_TimeLike_eom1}\\
\partial_+\partial_-{\bf Z}=&-\frac{\partial_+\varphi\partial_-{\bf Z}}{\tanh\varphi}-\frac{\partial_-\varphi\partial_+{\bf Z}}{\sinh\varphi\cosh\varphi}-\partial_-{\bf Z}^T\tilde{\mathcal{S}} \partial_+{\bf Z}\thickspace{\bf Z}.\label{AdS_TimeLike_eom2}
\end{align}
The complementary form of equation \eqref{AdS_TimeLike_eom2} reads
\begin{equation}
\partial_+\partial_-{\bf Z}=-\frac{\partial_+\varphi\partial_-{\bf Z}}{\sinh\varphi\cosh\varphi}-\frac{\partial_-\varphi\partial_+{\bf Z}}{\tanh\varphi}-\partial_-{\bf Z}^T\tilde{\mathcal{S}} \partial_+{\bf Z}\thickspace{\bf Z}.\label{AdS_TimeLike_eom3}
\end{equation}
Equation \eqref{AdS_TimeLike_eom1} and \eqref{AdS_TimeLike_eom2} appear in the exact same form in \cite{Grigoriev:2007bu,Rashkov:2008rm}. Regarding \cite{Miramontes:2008wt}, in the case of AdS$_2$ the equations obviously coincide, nevertheless this is not the case for AdS$_3$. The different form of the equations is due to the parametrization. 

Let us show how the different parametrizations are related in the case of \eqref{AdS_TimeLike_eom2}. We parametrize ${\bf Z}$ as ${\bf Z}=\begin{pmatrix} \cos u \\ \sin u\end{pmatrix}$, while $\tilde{\mathcal{S}}=I_2$. It is a matter of algebra to show that the equations of motion read
\begin{align}
\partial_+&\partial_-\varphi=\tanh\varphi\,\partial_-u \partial_+u+\frac{m_+m_-}{\Lambda^2}\sinh\varphi,\\
\partial_+&\left[\left(c+\textrm{sech}\varphi\right)\partial_-u\right]-\partial_-\left[\left(c+\cosh\varphi\right)\partial_+u\right]=0,
\end{align}
where $c$ is an free parameter. Implementing the non-local field redefinition 
\begin{equation}
\partial_-u=2\frac{\tanh^2\frac{\phi}{2}}{c+\textrm{sech}\varphi}\partial_-\theta,\qquad\partial_+u=-2\frac{\tanh^2\frac{\phi}{2}}{c+\cosh\varphi}\partial_+\theta
\end{equation}
and setting $c=-1$, the equations of motion assume the form
\begin{align}
\partial_+&\partial_-\varphi=\frac{\sinh\varphi}{\cosh^4\frac{\phi}{2}}\partial_-\theta \partial_+\theta+\frac{m_+m_-}{\Lambda^2}\sinh\varphi,\\
\partial_+&\left[\tanh^2\frac{\phi}{2}\,\partial_-\theta\right]+\partial_-\left[\tanh^2\frac{\phi}{2}\,\partial_+\theta\right]=0,
\end{align}
which is the one appearing in \cite{Miramontes:2008wt}. Finally, let us mention that these equations of motion follow from the Lagrangian density
\begin{equation}
\mathcal{L}=\frac{1}{4}\partial_+\varphi\partial_-\varphi+\tanh^2\frac{\phi}{2}\partial_-\theta\partial_+\theta+m_+m_-\sinh^2\frac{\varphi}{2}.
\end{equation}

Regarding the case of \eqref{AdS_TimeLike_eom3}, using the same parametrization and the non-local field redefinition
\begin{equation}
\partial_-u=-2\frac{\coth^2\frac{\phi}{2}}{c+\cosh\varphi}\partial_-\theta,\qquad\partial_+u=2\frac{\coth^2\frac{\phi}{2}}{c+\textrm{sech}\varphi}\partial_+\theta
\end{equation}
for $c=1$ it follows that the equations of motion assume the form
\begin{align}
\partial_+&\partial_-\varphi=-\frac{\sinh\varphi}{\sinh^4\frac{\phi}{2}}\partial_-\theta \partial_+\theta+\frac{m_+m_-}{\Lambda^2}\sinh\varphi,\\
\partial_+&\left[\coth^2\frac{\phi}{2}\,\partial_-\theta\right]+\partial_-\left[\coth^2\frac{\phi}{2}\,\partial_+\theta\right]=0.
\end{align}
These equations of motion follow from the Lagrangian density
\begin{equation}
\mathcal{L}=\frac{1}{4}\partial_+\varphi\partial_-\varphi+\coth^2\frac{\phi}{2}\partial_-\theta\partial_+\theta+m_+m_-\sinh^2\frac{\varphi}{2}.
\end{equation}

Finally, in order to point out the relation between the reduced systems corresponding to time-like and space-like reduction for any $n$, let us mention that for any $n$ the column ${\bf Z}$ can be parametrized as
\begin{equation}
{\bf Z}=\begin{pmatrix} \cos u \\ \sin u \,\hat{n}\end{pmatrix},
\end{equation}
where $\hat{n}$ is a $(n-1)\times 1$ column matrix of unit norm.

\subsection{AdS Light-like reduction \& Minkowski world-sheet}
In this section we analyze light-like reduction of AdS spaces. We remind the reader that this case corresponds to $t=0$, i.e. $\partial_\pm Y\cdot \partial_\pm Y=0$. It is customary to define the Pohlmeyer field via the equation
\begin{equation}
\partial_+Y\cdot\partial_-Y=m_+ m_-\exp \varphi.
\end{equation}
In this case, given two light-like vectors we can always define a time-like and a space-like one. Thus, $\mathcal{M}_3=\mathbb{R}^{(2,1)}$, which implies that $\mathcal{M}_{n-1}=\mathbb{R}^{(0,n-1)}$. It is a matter of algebra to show that
\begin{align}
\partial_\sigma Y\cdot\partial_\sigma Y&=+\frac{m_+m_-}{2}\exp\varphi,\\
\partial_\tau Y\cdot\partial_\tau Y&=-\frac{m_+m_-}{2}\exp\varphi,\\
\partial_\sigma Y\cdot\partial_\tau Y&=0.
\end{align}
Obviously, one of the vector $\partial_\sigma Y$ and $\partial_\tau Y$ is space-like, while the other vector one time-like, as required. 

Let is also comment on the reduced system. In the case of AdS$_3$, equation \eqref{eq:eom_light_1_final} is satisfied identically as ${\bf v}={\bf 0}$, and \eqref{eq:eom_light_2_final} reduces to sinh- or cosh-Gordon. In the degenerate case of ${\bf a}_-$  being null, which was omitted, the equation reduces to the Liouville equation and the configuration actually lives in AdS$_2$. For AdS$_4$ parametrizing $v=\cos\beta$ and ${\bf v}=(\sin\beta)$, the reduced system reads
\begin{align}
\partial_-\partial_+\beta&=\mu\exp\left(-\varphi\right)\sin\beta,\label{eq:eom_ads_4_a}\\
\partial_+\partial_-\varphi&=\mu\left[\cos\beta\exp\left(-\varphi\right)-s_{\mu}\exp\left(\varphi\right)\right],\label{eq:eom_ads_4_b}
\end{align}
which matches the results of \cite{Rashkov:2008rm}.
\subsection{AdS Space-like reduction \& Minkowski world-sheet}
In this section we analyze the space-like reduction of AdS spaces. We remind the reader that this case corresponds to $t=1$, i.e. $\partial_\pm Y\cdot \partial_\pm Y=m^2_\pm$. Contrary to the time-like case, we define the inner product as
\begin{equation}
\partial_+Y\cdot\partial_-Y=m_+ m_-\cos \varphi,
\end{equation}
it follows that $\mathcal{M}_3=\mathbb{R}^{(1,2)}$. It is a matter of algebra to show that
\begin{align}
\partial_\sigma Y\cdot\partial_\sigma Y&=\frac{m_+^2+m_-^2}{4}+\frac{m_+m_-}{2}\cos\varphi,\\
\partial_\tau Y\cdot\partial_\tau Y&=\frac{m_+^2+m_-^2}{4}-\frac{m_+m_-}{2}\cos\varphi,\\
\partial_\sigma Y\cdot\partial_\tau Y&=\frac{m_+^2-m_-^2}{4}.
\end{align}
Thus, both vectors $\partial_\sigma Y$ and $\partial_\tau Y$ are space-like as required.

In this case, it follows that $\mathcal{S}=I_{1,n-2}$, i.e. the Killing metric of $SO(1,n-2)$. As $f(\varphi)=\cos\varphi$, it is required that $\tilde{s}t=1$. For the space-like reduction $t=1$, thus $\tilde{s}=1$. As a result $\tilde{\mathcal{S}}=I_{1,n-1}$, which is the Killing metric of $SO(1,n-1)$. Again, the reduced theory is formulated as the $SO(1,n)/SO(1,n-1)$ perturbed WZW model, see section 5.1 of \cite{Miramontes:2008wt}. In order to make contact with the results of this work, once more we need to implement non-local field redefinitions.

For completeness, we also present the equation of the reduced theory, which read
\begin{align}
\partial_+\partial_-\varphi=&\tan\varphi\,\partial_-{\bf Z}^T\tilde{\mathcal{S}} \partial_+{\bf Z}+\frac{m_+m_-}{\Lambda^2}\sinh\varphi,\label{AdS_SpaceLike2_eom1}\\
\partial_+\partial_-{\bf Z}=&-\frac{\partial_+\varphi\partial_-{\bf Z}}{\tan\varphi}-\frac{\partial_-\varphi\partial_+{\bf Z}}{\sin\varphi\cos\varphi}-\left(\partial_-{\bf Z}^T\tilde{\mathcal{S}} \partial_+{\bf Z}\right){\bf Z}.\label{AdS_SpaceLike2_eom2}
\end{align}
The complementary form of equation \eqref{AdS_SpaceLike2_eom2} reads
\begin{equation}
\partial_+\partial_-{\bf Z}=-\frac{\partial_+\varphi\partial_-{\bf Z}}{\sin\varphi\cos\varphi}-\frac{\partial_-\varphi\partial_+{\bf Z}}{\tan\varphi}-\left(\partial_-{\bf Z}^T\tilde{\mathcal{S}} \partial_+{\bf Z}\right){\bf Z}.\label{AdS_SpaceLike2_eom3}
\end{equation}

We parametrize ${\bf Z}$ as ${\bf Z}=\begin{pmatrix} \hat{n}\cosh u\\ \sinh u\end{pmatrix}$, where $\hat{n}$ is a $(n-1)\times 1$ column matrix of unit norm. Consequently, the metric $\tilde{\mathcal{S}}$ reads $\tilde{\mathcal{S}}=\begin{pmatrix}
I_{n-1} & 0 \\ 0 & -1
\end{pmatrix}$. It is evident that equations \eqref{AdS_SpaceLike2_eom1}-\eqref{AdS_SpaceLike2_eom3} are obtained from equations \eqref{AdS_TimeLike_eom1}-\eqref{AdS_TimeLike_eom3} via the double analytic continuation $u\rightarrow iu+\pi/2$ and $\varphi\rightarrow i\varphi$.

There is also an alternative choice for the inner product. We can define it as
\begin{equation}
\partial_+Y\cdot\partial_-Y=m_+ m_-\cosh \varphi,
\end{equation}
so that $\mathcal{M}_3$ can be $\mathbb{R}^{(2,1)}$. It is a matter of algebra to show that
\begin{align}
\partial_\sigma Y\cdot\partial_\sigma Y&=\frac{m_+^2+m_-^2}{4}+\frac{m_+m_-}{2}\cosh\varphi,\\
\partial_\tau Y\cdot\partial_\tau Y&=\frac{m_+^2+m_-^2}{4}-\frac{m_+m_-}{2}\cosh\varphi,\\
\partial_\sigma Y\cdot\partial_\tau Y&=\frac{m_+^2-m_-^2}{4}.
\end{align}
The consistency of the basis requires
\begin{equation}
\cosh\varphi\geq \frac{m_+^2+m_-^2}{2\vert m_+ m_-\vert},
\end{equation}
so that one of the vectors $\partial_\sigma Y$ and $\partial_\tau Y$ is time-like and the other one space-like.

In this case, it follows that $\mathcal{S}=I_{n-1}$, i.e. the Killing metric of $SO(n-1)$. As $f(\varphi)=\cosh\varphi$, it is required that $\tilde{s}t=-1$. For the space-like reduction $t=1$, thus $\tilde{s}=-1$. As a result $\tilde{\mathcal{S}}=I_{1,n-1}$, which is the Killing metric of $SO(1,n-1)$. The reduced theory is formulated as the $SO(1,n)/SO(1,n-1)$ perturbed WZW model, see section 5.1 of \cite{Miramontes:2008wt}.

For completeness, we also present the equation of the reduced theory, which read
\begin{align}
\partial_+\partial_-\varphi=&-\tanh\varphi\,\partial_-{\bf Z}^T\tilde{\mathcal{S}} \partial_+{\bf Z}+\frac{m_+m_-}{\Lambda^2}\sinh\varphi,\label{AdS_SpaceLike1_eom1}\\
\partial_+\partial_-{\bf Z}=&-\frac{\partial_+\varphi\partial_-{\bf Z}}{\tanh\varphi}-\frac{\partial_-\varphi\partial_+{\bf Z}}{\sinh\varphi\cosh\varphi}+\partial_-{\bf Z}^T\tilde{\mathcal{S}} \partial_+{\bf Z}\thickspace{\bf Z}.\label{AdS_SpaceLike1_eom2}
\end{align}
The complementary form of equation \eqref{AdS_SpaceLike1_eom2} reads
\begin{equation}
\partial_+\partial_-{\bf Z}=-\frac{\partial_+\varphi\partial_-{\bf Z}}{\sinh\varphi\cosh\varphi}-\frac{\partial_-\varphi\partial_+{\bf Z}}{\tanh\varphi}+\partial_-{\bf Z}^T\tilde{\mathcal{S}} \partial_+{\bf Z}\thickspace{\bf Z}.\label{AdS_SpaceLike1_eom3}
\end{equation}
Notice that the sign of the terms involving $\partial_-{\bf Z}^T\tilde{\mathcal{S}} \partial_+{\bf Z}$ as well as the metric $\tilde{\mathcal{S}}$ are different with respect the time-like case. We parametrize ${\bf Z}$ as ${\bf Z}=\begin{pmatrix} \cosh u \\ \hat{n}\sinh u\end{pmatrix}$, where $\hat{n}$ is a $(n-1)\times 1$ column matrix of unit norm. Consequently, the metric $\tilde{\mathcal{S}}$ reads $\tilde{\mathcal{S}}=\begin{pmatrix}
-1 & 0 \\ 0 & I_{n-1}
\end{pmatrix}$. It is evident that equations \eqref{AdS_SpaceLike1_eom1}-\eqref{AdS_SpaceLike1_eom3} are obtained from equations \eqref{AdS_TimeLike_eom1}-\eqref{AdS_TimeLike_eom3} via the analytic continuation $u\rightarrow iu$.

\section{Details of the solution of the auxiliary system}\label{sec:details}
In this section we provide some details on the final part of the derivation of section \ref{eq:sec_solution}. To begin with the matrix $V$, which appears in \eqref{eq:Pohl_derivatives_matrix}, has the structure
\begin{equation}
V^T=\begin{pmatrix}\label{eq:V_matrix_struc}
v_1^T&\dots&v_{n+2}^T
\end{pmatrix},
\end{equation}
while the inverse matrix has the structure
\begin{equation}\label{eq:V_inv_matrix_struc}
V^{-1}=\begin{pmatrix}
v_1^\prime & \dots & v_{n+2}^\prime
\end{pmatrix},
\end{equation}
where, obviously, the vectors $v_\alpha^\prime$ obey $v_\alpha^T v_\beta^\prime=\delta_{\alpha\beta}$. On a similar token the currents $\hat{j}_\pm$ have the following form
\begin{align}
\hat{j}_+&=\frac{s}{\Lambda^2}\left[v_{n+1}v_{n+2}^T-v_{n+2}v_{n+1}^T\right],\\
\hat{j}_-&=\frac{s}{\Lambda^2}\left[v_{n}v_{n+2}^T-v_{n+2}v_{n}^T\right].
\end{align}
It is a matter of algebra to show that
\begin{equation}
V\hat{j}_+ V^{-1}=\begin{pmatrix}
{\bf 0}_{n-1} & {\bf 0} & {\bf 0} & {\bf 0}\\
{\bf 0}^T & 0 & 0 & s \frac{m_+ m_-}{\Lambda^2}f(\varphi)\\
{\bf 0}^T & 0 & 0 & s t \frac{m_+^2}{\Lambda^2}\\
{\bf 0}^T & 0 & -1 & 0
\end{pmatrix},
\end{equation}
\begin{equation}
V\hat{j}_- V^{-1}=\begin{pmatrix}
{\bf 0}_{n-1} & {\bf 0} & {\bf 0} & {\bf 0}\\
{\bf 0}^T & 0 & 0 & s t \frac{m_+^2}{\Lambda^2}\\
{\bf 0}^T & 0 & 0 & s \frac{m_+ m_-}{\Lambda^2}f(\varphi)\\
{\bf 0}^T & -1 & 0 & 0
\end{pmatrix},
\end{equation}
where ${\bf 0}_{n-1}$ is the $(n-1)\times(n-1)$ zero matrix and ${\bf 0}$ is the  $(n-1)\times1$ zero matrix.

The explicit form of the vectors $v_\alpha^\prime$ is
\begin{equation}\label{eq:inverse_mat_1}
v_i^\prime=s_i \eta v_i, \qquad v_{n+2}^\prime=\frac{s}{\Lambda^2} \eta v_{n+2}
\end{equation}
where $i=1,\dots,n-1$ and $\eta$ is the metric of the enhanced space. The rest of the columns read
\begin{align}
v^\prime_n&=\frac{1}{m_-^2\left(t^2-f^2\right)}\eta\left[t v_n-\frac{m_-}{m_+}f v_{n+1}\right],\label{eq:inverse_mat_2}\\
v^\prime_{n+1}&=\frac{1}{m_+^2\left(t^2-f^2\right)}\eta\left[t v_{n+1}-\frac{m_+}{m_-}f v_{n}\right].\label{eq:inverse_mat_3}
\end{align}


%

\end{document}